\documentclass[aps,prl,twocolumn,showpacs,superscriptaddress,preprintnumbers,amsmath,amssymb,longbibliography]{revtex4-1}

\newcommand{\vectau}{\boldsymbol{\tau}}
\newcommand{\veck}{\mathbf{k}}

\newcommand{\veca}{{\bf a}}
\newcommand{\vecA}{{\bf A}}

\newcommand{\vecv}{{\bf v}}

\usepackage[pdftex,colorlinks=true,pdfstartview=FitV,linkcolor=blue,citecolor=blue,urlcolor=blue]{hyperref}

\usepackage{xcolor}
\usepackage{soul}
\usepackage{hyperref}
\usepackage{graphicx}
\usepackage{bm}
\usepackage{lineno}
\usepackage{multirow}
\usepackage{footnote}
\usepackage{soul}
\usepackage[colorinlistoftodos]{todonotes}

\begin{document}

\title{Structural classification of boron nitride twisted bilayers and ab initio investigation of their stacking-dependent electronic structure}

\author{Sylvain Latil}
\affiliation{Universit\'e Paris-Saclay, CEA, CNRS, SPEC, 91191 Gif-sur-Yvette, France}

\author{Hakim Amara}
\affiliation{Universit\'e Paris-Saclay, ONERA, CNRS, Laboratoire d'\'{e}tude des microstructures (LEM), 92322, Ch\^atillon, France}
\affiliation{Universit\'e de Paris, Laboratoire Mat\'eriaux et Ph\'enom\`enes Quantiques (MPQ), CNRS-UMR7162, 75013 Paris, France}

\author{Lorenzo Sponza}
\affiliation{Universit\'e Paris-Saclay, ONERA, CNRS, Laboratoire d'\'{e}tude des microstructures (LEM), 92322, Ch\^atillon, France}
\affiliation{European Theoretical Spectroscopy Facility (ETSF), B-4000 Sart Tilman, Li\`ege, Belgium}

\begin{abstract}
Since the discovery of superconductive twisted bilayer graphene which initiated the field of twistronics, moir\'e systems have not ceased to exhibit fascinating properties.
We demonstrate that in boron nitride twisted bilayers, for a given moir\'e periodicity, there are five different stackings which preserve the monolayer hexagonal symmetry (i.e. the invariance upon rotations of 120$^\circ$) and not only two as always discussed in literature.
We introduce some definitions and a nomenclature that identify unambiguously the twist angle and the stacking sequence of any hexagonal bilayer with order-3 rotation symmetry.
Moreover, we employ density functional theory to study the evolution of the band structure as a function of the twist angle for each of the five stacking sequences of boron nitride bilayers.
We show that the gap is indirect at any angle and in any stacking, and identify features that are conserved within the same stacking sequence irrespective of the angle of twist.
\end{abstract}

\date{\today}

\maketitle

Initiated by twisted bilayer graphene, moir\'e systems formed of 2D atomic layers have recently been established as a unique playground for highlighting novel and fascinating properties~\cite{Carr2020}.
A tiny twist between the two van der Waals atomic layers can modify deeply their electronic properties as a consequence of the flattening of the band dispersion.
In graphene, a flat moir\'e mini-band appears at specific ``magic angles"~\cite{Suarez2010, Bistritzer2011} whose occupation drives superconductive/insulating transitions which open new perspectives on the investigation of strong correlation in 2D systems~\cite{Balents2020, Kennes2021, Liu2021}.
In gapped twisted bilayers (e.g. semiconducting transition metal dichalcogenides) the moir\'e bands have an impact on the optical properties. For instance, by varying the twist angle it is possible to modulate the exciton lifetime~\cite{Choi2021}, or the energy and intensity of emitted light~\cite{Alexeev2019a,Jin2019,Seyler2019,Tran2019}. In these systems, flat bands give rise to intriguing phenomena without the need of being twisted by specific ``magic" angles~\cite{Wu2019, Ruiz2019, Ochoa2020}.

Hexagonal boron nitride (hBN) is a cardinal compound in 2D materials research.
Used mostly as incapsulating layer, it has nonetheless attracting properties on its own respect, mainly because of its large band gap ($>$~6~eV)~\cite{Blase1995, Galvani2016} which is at the origin of a strong UV emission~\cite{Schue2019, Rousseau2021}, single photon emission~\cite{Tran2016b, Bourrellier2016, Grosso2017a, Li2019a, Fischer2021, Libbi2021} and its application as gating layer in 2D electronics~\cite{Britnell2012, Levendorf2012a, Jang2016, Knobloch2021}.
Recently ferroelectricity has been enabled in twisted hBN bilayers, thus expanding further its range of applications~\cite{Yasuda2021, Woods2021}.
In the bulk phase and in thin layers its optical properties are driven by excitons~\cite{Paleari2018}.
In hBN moir\'e systems, Lee and coworkers~\cite{Lee2021} observed an increase of the luminescence intensity and a decrease of the sub-band gapwidth for increasing twist angles.
From the standpoint of atomistic simulations, geometries with small rotation angles require very large periodic cells (order of thousands of atoms) which are out of reach for most self-consistent numerical approaches~\cite{Zhao2020}.
As for graphene~\cite{Suarez2010, Trambly2010}, tight-binding or continuous models based on the $k \cdot p$ approximation are more adapted to deal with very large systems and have therefore been developed~\cite{Xian2019, Ochoa2020, Liu2021, Zhao2021}.
However these studies are incomplete on two aspects.
First, the very nature of the band gap is still not elucidated while it obviously rules the optical and excitonic properties of monolayer and bulk hBN~\cite{Blase1995, Wirtz2006, Galvani2016}.
Second, the stacking sequence in bilayers is seldom considered and, when it has been, only two geometries were taken into account~\cite{Zhao2020, Zhao2021}.
Yet, it has been shown that the stacking sequence strongly influences the character of the gap~\cite{Sponza2018, Paleari2018, Henriques2022} through long range interplanar interactions.

In this Letter, we investigate the electronic structure of twisted hBN bilayers by taking into account fully and on the same footing its dependence on the twist angle and the stacking sequence.
As a first step, we demonstrate the existence of five and only five different stacking possibilities to construct hBN bilayers with hexagonal symmetry and provide a non-ambiguous nomenclature applicable to untwisted configurations as well and to any other homobilayer formed of hexagonal 2D materials.
Stemming from this symmetry analysis, we employed density functional theory (DFT) to investigate the evolution of the band structure as a function of the twist angle for each of the five stackings.

\section{Geometrical analysis}

\begin{figure*}
  \includegraphics[width=.99\linewidth]{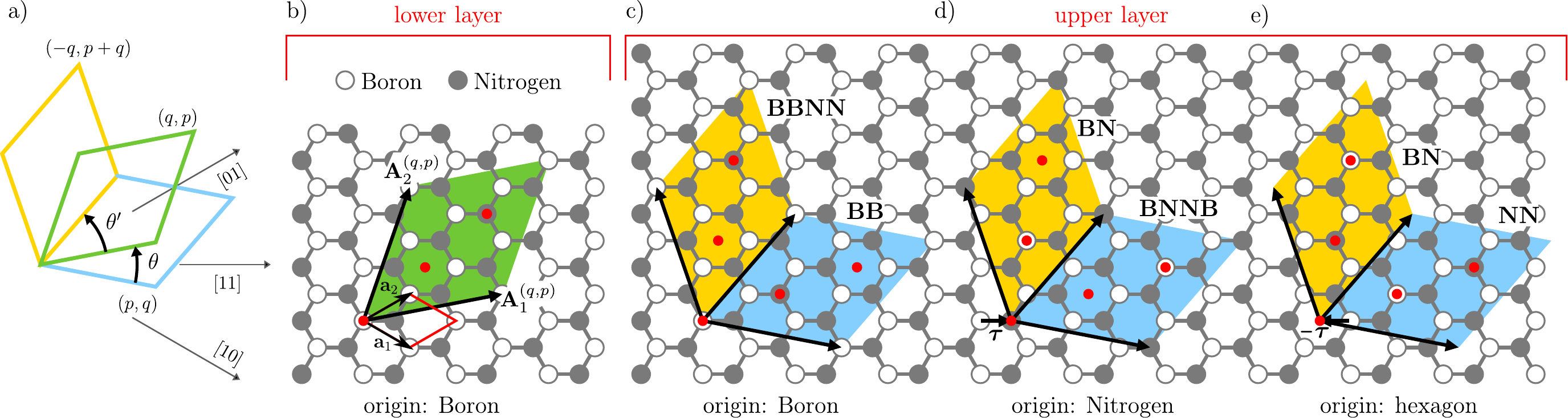}
  \caption{a) Graphical representation of $\theta$ and $\theta^\prime$
    angles according to the $\{p,q\}$ integers.
    b) The lower layer supercell is $(q,p)_\text{B}$. c) d) and e) The   supercells of the upper layer $(p,q)_\text{X}$
    with X = B, N or H respectively are drawn in blue, and the corresponding $(-q,p+q)_\text{X}$ supercells in yellow.
    High symmetry points are reported as red dots. 
     In the examples $p=2$ and $q=1$.}  
  \label{fig:construction}
\end{figure*}

To construct a tiling of rotated bilayers preserving long range translational symmetries, we first define coincident supercells~
\cite{Kolmogorov2006,Mele2010}.
Let us take a honeycomb lattice with primitive vectors \({\bf a}_1\) and
\({\bf a}_2\) forming an angle of {60$^\circ$} and with  the two atoms of the cell separated by $\bm{\tau}$.
Then we define the $(q,p)$ hexagonal supercell as resulting from the vectors ${\bf A}^{(q,p)}_i = \sum_j M^{(q,p)}_{ij}{\bf{a}}_j$
defined by means of the matrix
\begin{equation}
  \mathrm{M}^{(q,p)} = 
  \begin{bmatrix}
    q & p \\ -p & p+q
  \end{bmatrix}\,.
  \label{eq:matrix}
\end{equation}
Similarly we use equation~\eqref{eq:matrix} to introduce the $(p,q)$ and the $(-q,p+q)$ supercells.
The resulting twist angles are given respectively by the formulae:
\begin{equation*}
  \label{equ:tantheta}
  \tan\theta= \frac{\sqrt{3}(p^2-q^2)}{p^2+q^2+4pq} 
  \; \text{ or } \;
  \tan\theta^\prime= \frac{\sqrt{3}( q^2+2pq ) }{2p^2-q^2+2pq}. 
\end{equation*}
The supercells defined above and the resulting twist angles
are sketched in Figure~\ref{fig:construction}.a.

The $p$ and $q$
integers obey to some constraints: they
must be different and non zero, otherwise they lead to twist angles of 0$^\circ$ or 60$^\circ$, they must have no common divisor,
and the case
$p-q$ multiple of 3 has to be excluded as it corresponds to non-primitive moir\'e supercells.
Moreover, since twist angles 
are defined modulo 60$^\circ$, the definition of the $\mathrm{M}^{(\alpha,\beta)}$ matrix is not unique.
We will then restrict ourselves arbitrarily to cases
$p>q$ which imply that
angles are positive and
$\theta+\theta'=60^\circ$.
Note finally that the notation introduced here for twisted bilayers can be employed also for untwisted structures taking $q=0$ and $p=1$.
Despite these constraints and arbitrary choices, we demonstrate in Appendix E that all twisted angles corresponding to periodic moiré patterns can be expressed through an appropriate choice of the $(q,p)$ pair.

\begin{table}[b]
  \begin{tabular}{ c c l c c }
    \hline
  \hline
    upper & twist & symm. & stacking & double \\
    layer & angle & group & sequence & coincidence\\
    \hline
    $(p,q)_B$ &
    $+\theta$ &
    $68\,(p321)$ &
    \textbf{BB} &
    no\\
    $(p,q)_N$ &
    $-\theta'$ &
    $68\,(p321)$ &
    \textbf{BNNB} &
    yes\\
    $(p,q)_H$ &
    $+\theta$ &
    $68\,(p321)$ &
    \textbf{NN} &
    no\\
    $(-q,p+q)_B$ &
    $-\theta'$ &
    $67\,(p312)$ &
    \textbf{BBNN} &
    yes\\
    $(-q,p+q)_N$ &
    $+\theta$ &
    $65\,(p3)$ &
    \textbf{BN} &
    no \\
    $(-q,p+q)_H$ &
    $+\theta$ &
    $65\,(p3)$ &
    \textbf{BN} &
    no \\
   \hline \hline
  \end{tabular}
  \caption{\label{tab:list} The geometry of
    the five stackings of hBN twisted bilayers. The lower layer is based on the
    $(q,p)_B$ supercell.}
\end{table}

\begin{figure}
  \includegraphics[width=.99\linewidth]{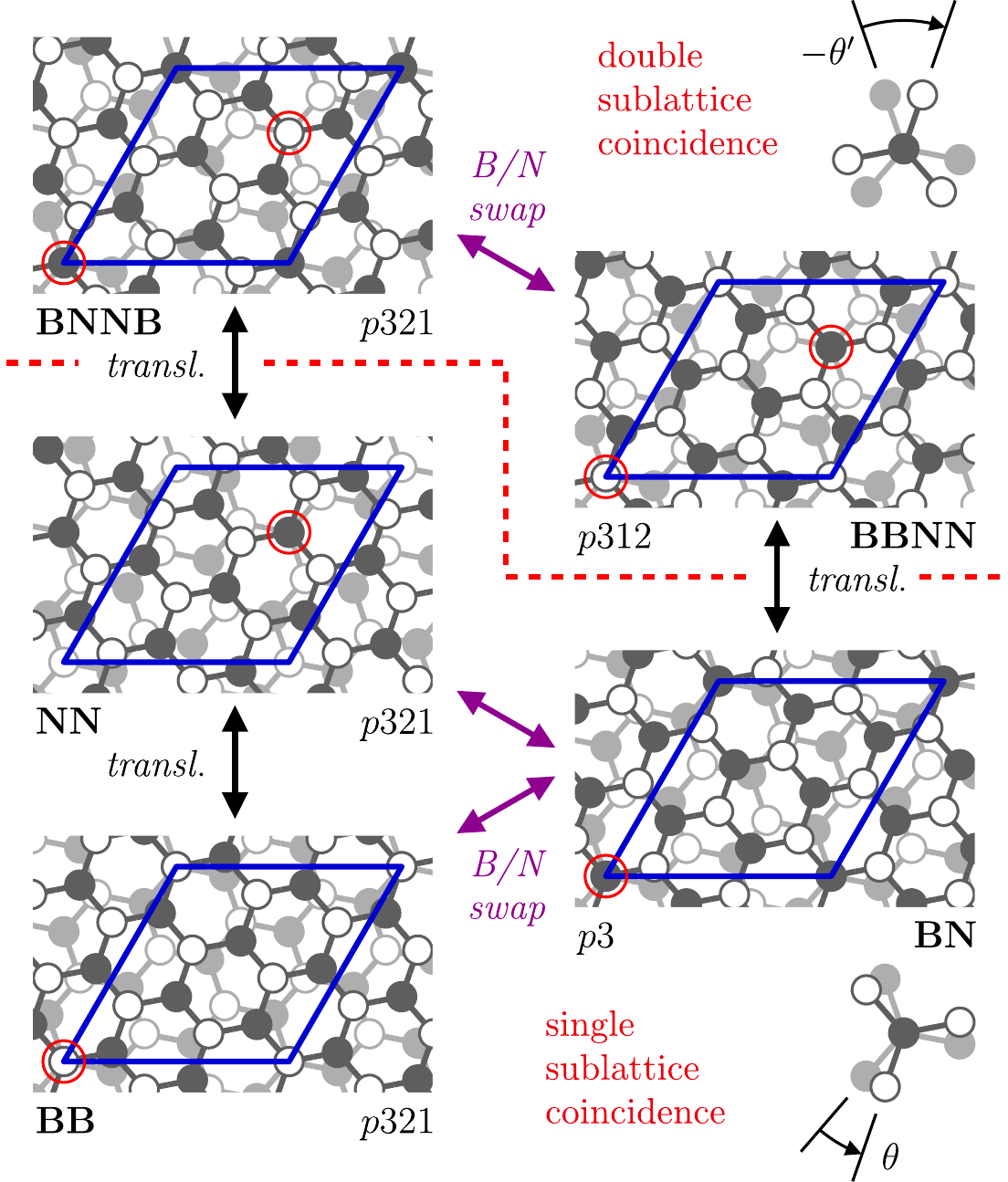}
  \caption{The five stackings of hBN moir\'{e} structures,
    with $p=2$ and $q=1$.
    The sublattice coincidences are highlighted with red circles.
    The reported symmetry groups are for twisted bilayers (untwisted bilayers have different symmetry properties).
    \label{fig:five}}
\end{figure}

Stacking the correct supercells is not enough to construct moir\'e hexagonal bilayers because the respective alignment is also crucial.
Let us introduce a subscript labelling the origin of the supercell (B~=~boron, N~=~nitrogen, H~=~hexagon center).
Without loss of generality we will always consider the supercell of the lower layer as being $(q,p)_\text{B}$ (cfr. Figure~\ref{fig:construction}b) while that of the upper layer can be any of $(p,q)_\text{B,N,H}$ or $(-q,p+q)_\text{B,N,H}$.
As a consequence, one ends up with six bilayers listed in Table~\ref{tab:list} and sketched in panels c), d) and e) of Figure~\ref{fig:construction} for the case $p=2, q=1$.
In each supercell there are three direct-space high-symmetry points (red bullets in Figure~\ref{fig:construction}): the points $(0\;\;0)$, $(1/3\;\;1/3)$ and $(2/3\;\,2/3)$ in the supercell reduced coordinates.
Depending on the coincident atoms in these points, one can
distinguish between
(i)~two geometries with a double sublattice coincidence per cell,
the $(p,q)_\text{N}$ and the $(-q,p+q)_\text{B}$ ones, with a twist angle $-\theta'$,
and (ii)~the remaining four geometries with a single sublattice coincidence per cell and an angle of twist $\theta$.
However it is trivial to demonstrate that the bilayers resulting from the stacking of $(-q,p+q)_{\text{N,H}}$ on the $(q,p)_\text{B}$ are related by a simple
inversion and are therefore identical.
All this boils down to five
hexagonal stackings for the generic twisted hBN bilayer.
It is important to stress that these stacking sequences are not related to the moir\'e periodicity and do not impose any constraint on the mutual orientation of the two layers. The orientation and the stacking are two independent degrees of freedom in the design of the bilayer structure.
As a consequence, we will designate univocally a twisted bilayer by the notation \emph{STACK}$(q,p)$ where 
the $\{p,q\}$ pair relates to the supercell and hence the moiré
periodicity and angles, and \emph{STACK}~=~BBNN, BNNB, BB, BN or NN relates to the atoms in the coincident sites.
Images of
these stackings, their layer symmetry group and
the transformations to
be applied to the upper layer to switch from one stacking to another
(swapping of B/N atoms
or translation by $\pm\boldsymbol{\tau}$) are summarized
in Figure~\ref{fig:five} and Table~\ref{tab:list}. 
It is worth recalling that with our conventions the angles are positive.
Their sign comes from the chirality of twisted bilayers and is defined according to the screw angle separating B-N bonds at the atom-on-atom coincidence sites of the supercell, as depicted in the insets of the Figure~\ref{fig:five}.

From this analysis, it appears that the five possible stackings of untwisted bilayers (and bulk) reported in literature~\cite{Mengle2019a, Gilbert2019} are actually special cases of a more general scheme, in agreement with the fact that the mutual orientation and the stacking sequence are two independent degrees of freedom.
However it is also important to stress that some ``special" structures, typically the untwisted bilayers, have higher symmetries than those reported in Table~\ref{tab:list} and Figure~\ref{fig:five}.

For comparison, in the case of graphene bilayers both B and N labels 
become C, so the possible stackings are only two, but they have higher symmetry.
The first belongs to the $p321$ layer group and to the odd bilayer graphene (BLG) set~\cite{Mele2010,Trambly2010,TramblyDeLaissardiere2012,Shallcross2010},
has a
single sublattice vertical coincidence per cell and the twist
angle is \(\theta\).
The second belongs to the
$p622$ layer group and to the even BLG set with hexagon-on-hexagon
or double sublattice coincidence. Its rotation angle is
\(-\theta^\prime\).
Finally, if we swap the values of $p$ and $q$,
we will obtain five new stackings
which are the mirror images of the pristine structures.
They will have the same electronic structure,
and the twist angles will be $+\theta'$ for the BNNB and BBNN 
and $-\theta$ for the BB, BN and  NN stackings.
Complete definitions and demonstrations are given in Appendix.

Before concluding this part, let us stress once more that the analysis we conducted on hBN bilayers is very general and easily applicable to any bilayer with hexagonal symmetry such as dichalcogenide bilayers or BN/graphene heterostructures, for instance.

\section{Electronic structure}

Based on our robust symmetry analysis, we clearly identify five different stackings of hBN bilayers.
Zhao and coworkers~\cite{Zhao2020} studied two of them (the NN and the BN one) with a DFT method based on a tight-binding Hamiltonian and demonstrated that the stacking sequence has an impact on the spatial localization of the top valence and bottom conduction states.
On the other hand, in a previous work~\cite{Sponza2018} we proved that interlayer coupling, and so the stacking, is crucial in the formation of the indirect band gap of the bulk phase.
These elements clearly indicate that a complete investigation involving all the stackings is mandatory.
As a consequence, 
we have performed first-principle simulations with density functional theory (DFT) to investigate the impact of the stacking sequence on the band gap.
We scrutinized thirty bilayers: six $\{p,q\}$ pairs per each stacking.
All the pertinent calculation parameters can be found in Appendix G.

As a first step, we investigated the structural stability of the five principal untwisted bilayers and identified two main groups (see Figure~11 in Appendix H).
In the three most stable structures (BN(0,1), BNNB(0,1) and BB(0,1)) the layers are separated by about 3.1~\AA.
The two least stable bilayers (BBNN(0,1) and NN(0,1)) are around 20 meV per formula unit at higher energy with larger equilibrium interlayer distances (around 3.4~\AA).
Regarding the electronic properties, untwisted bilayers with a boron-on-boron conicidence (BBNN(0,1) and BB(0,1)) have an indirect band gap whereas the other structures have a direct gap.
More details about the untwisted bilayers can be found in Appendix H.

\begin{figure}[t]
\centering
\includegraphics[width = 0.95\linewidth]{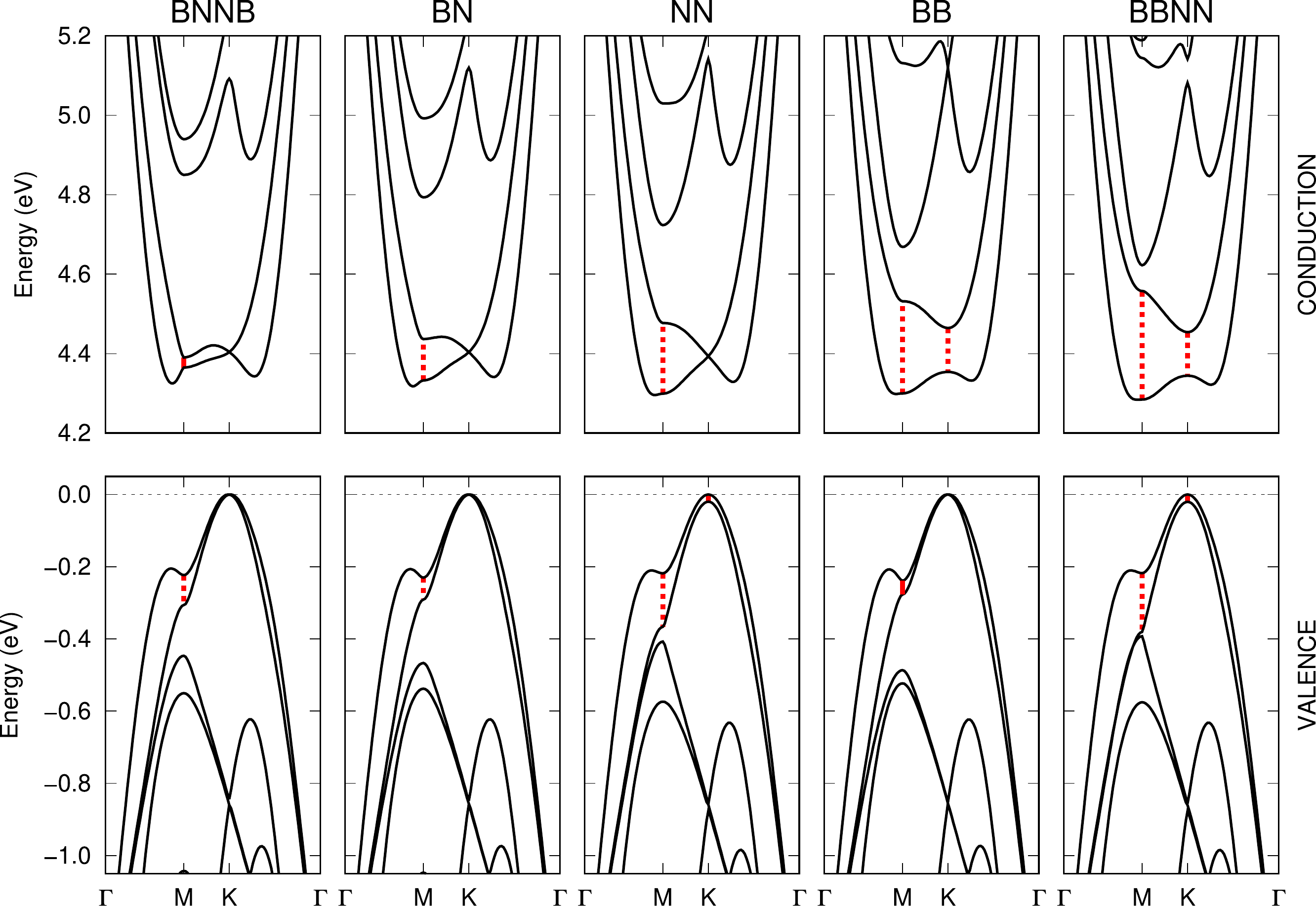}
\caption{Bottom conduction and top valence of the five principal stackings in the $(1,2)$ supercell. Red vertical dashed lines highlight the notable splittings at M and K reported also in Table~\ref{tab:bandstructure_per_stacking}.}
\label{fig:bands_vs_stacking}
\end{figure}

\begin{table}[b]
\centering
\begin{tabular}{c | c  c | c  c }
\hline \hline
\multirow{2}{*}{Structure} & \multicolumn{2}{ | c  | }{Top valence} & \multicolumn{2}{ | c  }{Bottom conduction} \\ 
     & @M & @K & @M & @K \\
	\hline
     BNNB(1,2) & 83 & - & 25 & -  \\
     BN(1,2)      & 61 & - & 104 & - \\
     NN(1,2)      & 148 & 20 & 178 & - \\
     BB(1,2)      & 38 & - & 232 & 110 \\
     BBNN(1,2) & 163 & 20 & 273 & 110 \\
\hline \hline
\end{tabular}
\caption{The band splitting (meV) at M and K in the top valence and bottom conduction of the (1,2) supercells. The symbol `-' indicates a band crossing.
 These features are highlighted with red vertical lines in Figure~\ref{fig:bands_vs_stacking}.  }
\label{tab:bandstructure_per_stacking}
\end{table}

\begin{table*}
\centering
\begin{tabular}{c c     c c c c c}
\hline \hline
family & $(q,p)$ cell & BNNB  & BN & NN & BB & BBNN \\
\hline 
\multirow{4}{*}{$\delta = 1$}  & (1,2) &   4.325~(71)    &  4.318~(76)   & 4.296~(88)     & 4.299~(~55)& 4.284~(60)\\ 
& (2,3) &   4.221~(30)   & 4.217~(34)    &  4.211~(38)    & 4.203~(~41) & 4.202~(42)\\
& (3,4) &      4.153~(15)  &  4.153~(16)  &  4.151~(17)    & 4.145~(~18) & 4.146~(19)\\ 
& (4,5) &     4.102~(~5)  & 4.103~(~5)   & 4.101~(~5)      & 4.098~(~~5)  & 4.099~(~5) \\ 
\hline 
\multirow{2}{*}{$\delta = 2$} & (1,3) &  4.284~(137)   & 4.284~(137)  &  4.284~(137)  & 4.284~(136) & 4.284~(136) \\ 
& (3,5) &  4.240~(~72)    & 4.241~(~72)  & 4.240~(~72)  & 4.240~(~72)  & 4.241~(~72) \\ 
\hline \hline
\end{tabular}
\caption{The DFT energy (eV) of the indirect band gap at different twist angles and stacking sequences. In parenthesis: energy difference between the direct and the indirect band gap in meV.}
\label{tab:gaps}
\end{table*}

We now discuss twisted bilayers.
We focus on the $(1,2)$ configuration for all stackings because notable effects are more distinguishable.
The DFT results are reported in Figure~\ref{fig:bands_vs_stacking} inside the Brillouin zone of the supercell.
It is important to recall that the preservation of the hexagonal symmetry of the
  supercell implies the conservation of their 
  order-3 rotation axes
  without which the equivalence between the $K$ points of the Brillouin zone would be lost.
 Interestingly, our calculations reveal that the gap is always indirect irrespective of the stacking with values around 4.3~eV (see first row of Table~\ref{tab:gaps}).
Indirect band gap has been reported also in untwisted bilayers of different stacking~\cite{Mengle2019a, Gilbert2019}, hBN multilayers~\cite{Paleari2018}, and bulk phases of different stackings~\cite{Sponza2018, Mengle2019a}, indicating that this is indeed a robust characteristic of BN multilayers. 
 By analyzing in details the electronic structure, we can distinguish the stackings according to characteristics at the $K$ and $M$ points.
In the valence region we observe that when N atoms are on top of each other (the NN and the BBNN stackings), a band crossing is avoided in the top valence at $K$ while the splitting between the HOMO and HOMO-1 at $M$ is the largest.
On the conduction band, the splitting between the LUMO and the LUMO+1 at $M$ is reduced along the sequence BBNN, BB, NN, BN and BNNB while the presence of B atoms on top of each other (BB and BBNN stackings) prevents a band crossing at $K$.
All the features discussed here are highlighted with dashed vertical red lines in Figure~\ref{fig:bands_vs_stacking} and reported in Table~\ref{tab:bandstructure_per_stacking}.
We expect these effects to be less important at extremal twist angles (i.e. close to 0$^\circ$ and 60$^\circ$) because the immediate surroundings of each atom change progressively.

\begin{figure}[b]
\centering
\includegraphics[width = 0.90\linewidth]{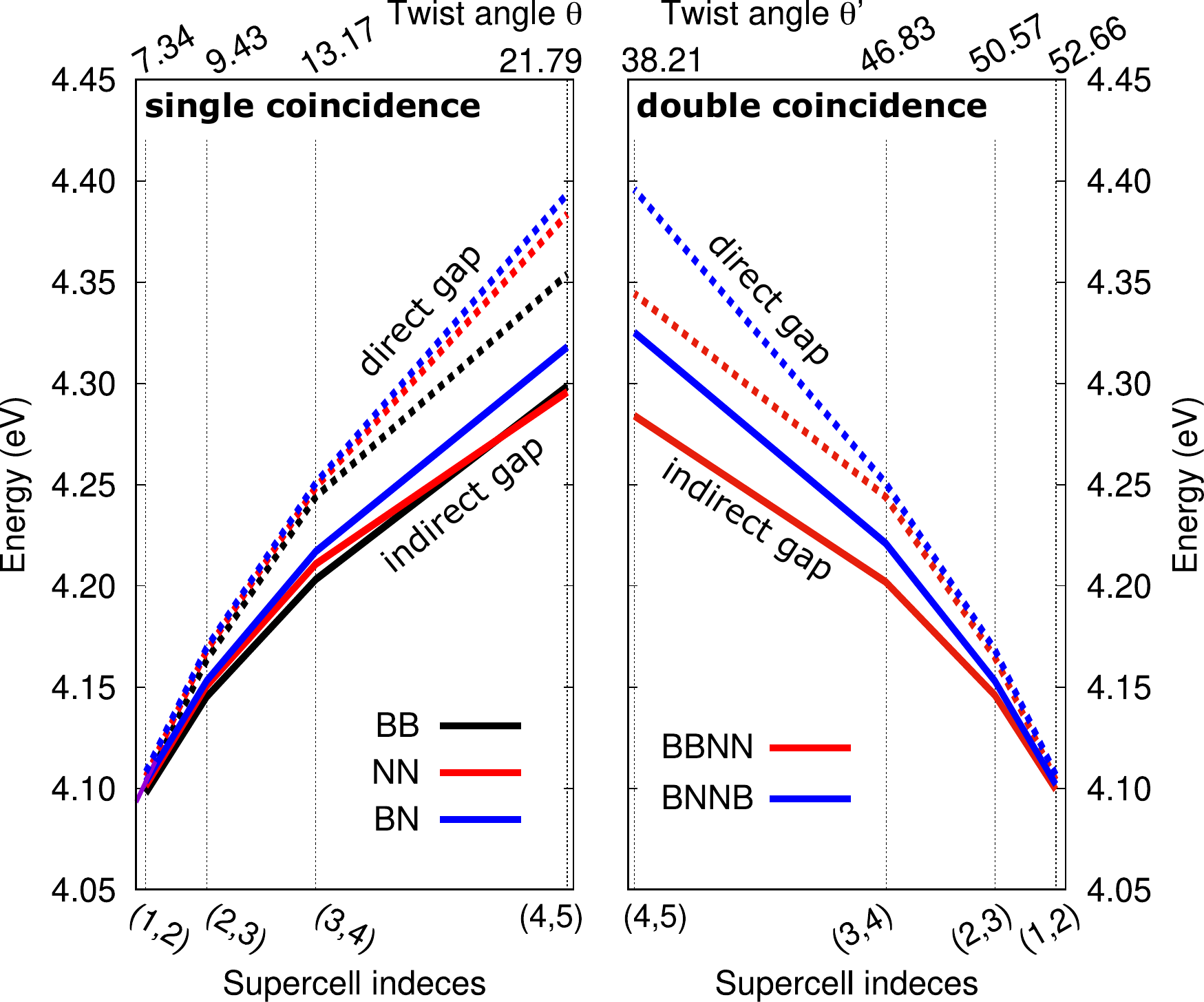}
\caption{Indirect gap (solid lines) and direct gap (dashed lines) of the five stackings as a function of the twist angle ($\theta$ or $\theta'$ depending on the stacking) within the $\delta = 1$ family.}
\label{fig:gap_vs_theta}
\end{figure}

Let us now discuss the evolution of the band gap as a function of the twist angle.
In Table~\ref{tab:gaps} and in Figure~\ref{fig:gap_vs_theta} we summarize our DFT results on the indirect band gap and the difference between direct and indirect gap.
First, we observe that the gapwidth gets smaller (higher) for smaller $\theta$ ($\theta'$), demonstrating a trend opposite to what predicted by continuous models \cite{Lee2021}.
Typically, for $\theta$ varying from 21.79$^\circ$~to 7.34$^\circ$, the gap decreases by about 5\%.
Secondly we observe that in each stacking the gap remains indirect at all angles.
This finding contrasts with density-functional tight-binding results where direct gaps at all twist angles are obtained instead~\cite{Zhao2020}.
A possible explanation of this discrepancy resides in a bad accounting of the interlayer interactions.
In fact, our study of the exciton dispersion~\cite{Sponza2018} demonstrates that the interlayer hopping terms are of paramount importance for the formation of the indirect band gap.
A more detailed analysis reported in Appendix I allows us to affirm that it is not an artifact coming from $\sigma$ or nearly-free-electron states located at higher energies~\cite{Posternak1983, Posternak1984,Blase1994a,Blase1994b,Blase1995,Hu2010,Paleari2019}.
We should stress that these results are reliable as long as one considers energy differences and trends, absolute gap energies being systematically underestimated by DFT.
Indeed, we expect hybrid potential calculations or quasiparticle corrections included via the GW approximation to be almost identical form one system to the other, and to have minor effect on the dispersion of $s$ and $p$ states~\cite{Blase1995, Paleari2018}. We are confident on this because of the successful use of the scissor operator in BN compounds~\cite{Galvani2016, Sponza2018, Henriques2022} and direct comparisons between different methods~\cite{Wirtz2006,Berseneva2013,Galvani2016, Paleari2018,Mengle2019a}. 

\begin{figure}[b]
\centering
\includegraphics[width = 0.99\linewidth]{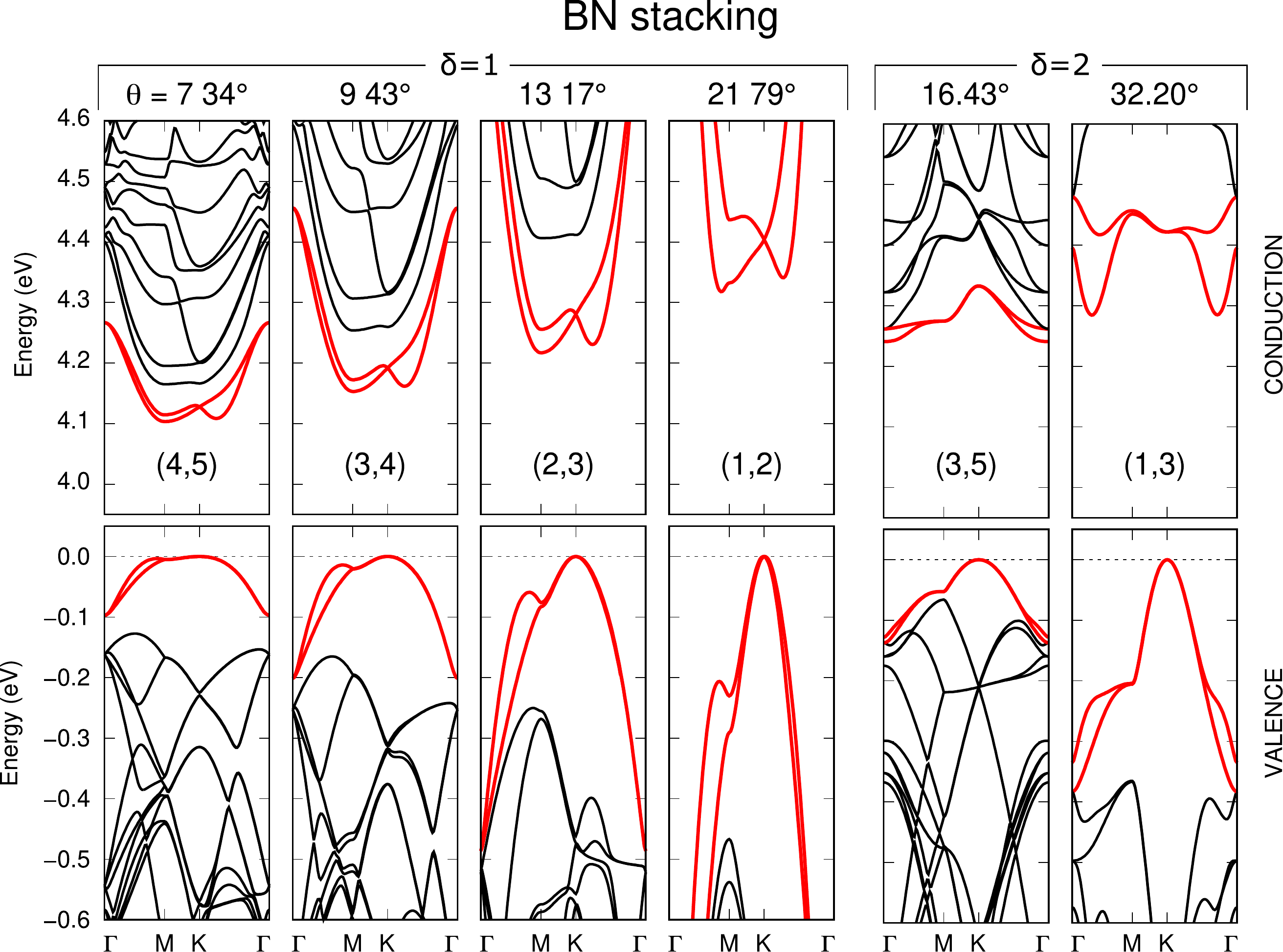}

\vspace{3mm}

\includegraphics[width = 0.99\linewidth]{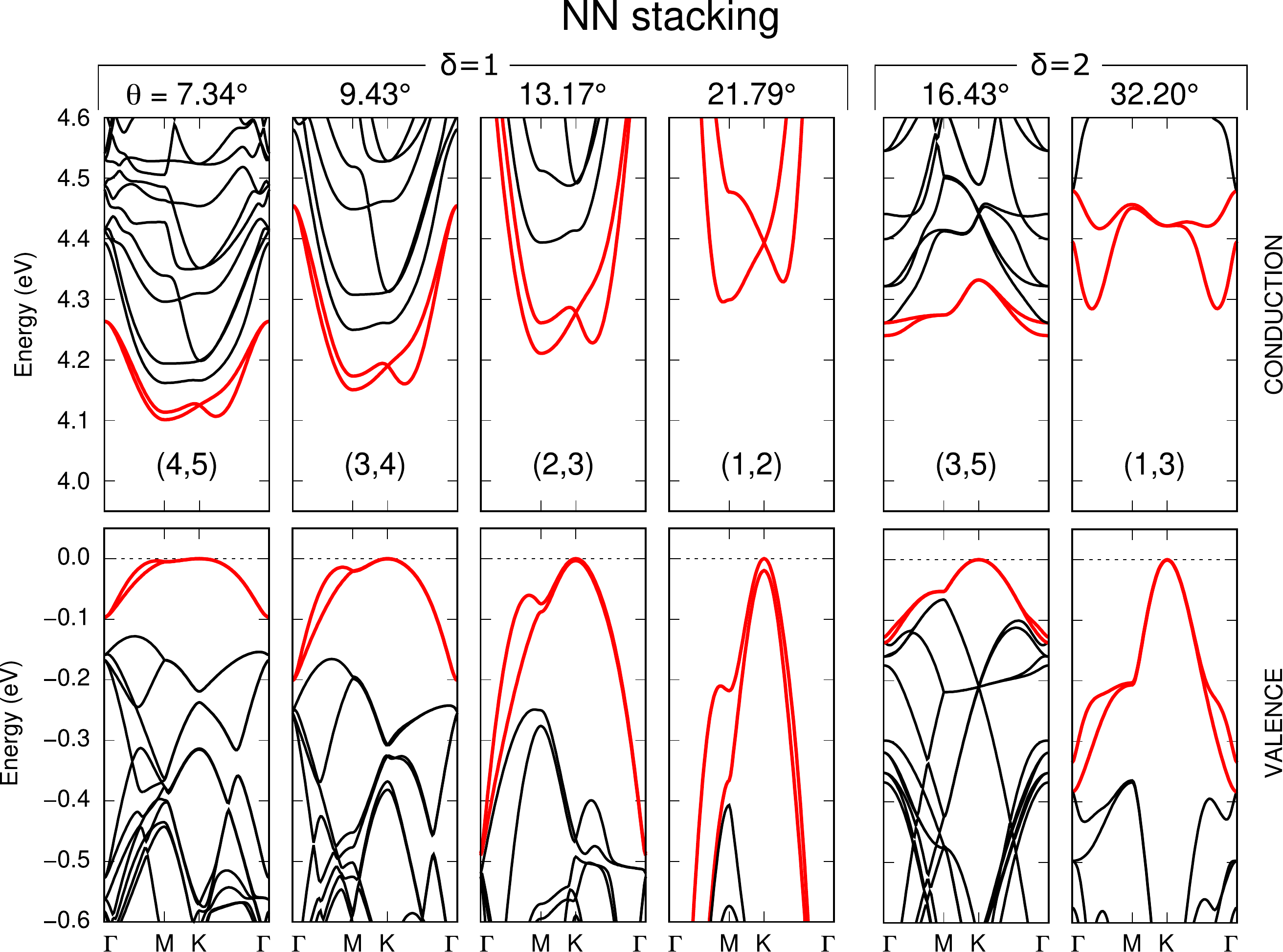}
\caption{Bottom conduction and top valence of the BN (top panel) and NN (bottom panel) stackings at different twist angles.}
\label{fig:bands_vs_angles}
\end{figure}

We can now pass to the investigation of the evolution of the full band structure as a function of the twist angle.
In the main text we discuss two paradigmatic stackings, the BN and the NN and we report the corresponding twelve band structure plots in Figure~\ref{fig:bands_vs_angles}.
We refer the reader to the Appendix K for the other bandplots.
We observe that conduction and valence bands get flatter at smaller $\theta$ (and larger $\theta^\prime$) as highlighted in Figure~\ref{fig:bands_vs_angles}. 
This implies the progressive creation of localized valence and conduction states in agreement with what shown by Zhao and coworkers~\cite{Zhao2020}.
For example, in the BN stacking at $\theta=$7.34$^\circ$, the HOMO and LUMO states are characterized by bandwidths around 0.09~eV and 0.16~eV, respectively.
Flatter bands are not observed since this would demand much smaller angles which are inaccessible with our numerical resources.
Because of the flattening of the bands, it is possible to tune the difference between indirect and direct gap through the twist angle, and so possibly to convert progressively the radiative decay pathway from a phonon-assisted emission to a direct recombination.  This may have strong impact on the intensity of emitted light (probability of recombination), its temperature dependence (through the coupling with phonons) and finally the life time of excitations. 

In addition we observe that $\{p,q\}$ pairs can be grouped into families defined by the parameter $\delta = | p - q |$ that characterizes the interplay between crystalline structure (twist angle) and electronic structure (bands).
In fact, the bands around the gap within the same family look similar but shrunk and flattened at small $\theta$ (or larger $\theta^\prime$).
Once more, the case $\delta$ multiple of 3 shall be excluded.
Consider the family $\delta = 1$, corresponding to the first four plots from the left in the band plots of Figure~\ref{fig:bands_vs_angles}.
Here the valence bands present a maximum in $K$ and are formed of two bands dispersing almost parabolically, up to $M$ where one of the two deviates with a small bump.
In conduction, two valleys are well discernible between $K$ and $\Gamma$ and around $M$, the latter forming the conduction band minimum.
Formally, the untwisted bilayers fall in this very family. However, their band structure presents unique characteristics and thus differs from that of the twisted counterparts (cfr the untwisted band structures reported in~\cite{Mengle2019a}).
At the time being, we can not identify the reason for this deviation, but we hypothesize that it is due to the fact that untwisted bilayers have higher symmetries than the twisted bilayers of the same stacking.
Because of this unicity, we have not added the untwisted band structures to the bandplots of Figures~\ref{fig:bands_vs_angles} and~\ref{fig:band_vs_angles_bis}. 

The last two plots from the left in the band plots of Figure~\ref{fig:bands_vs_angles} belong to the $\delta  = 2$ family. 
These bandplots look very different from those of the other family, even though the gap remains indirect with the top valence at $K$. 
As before, one can see common features within this family despite the band shrinking.
The valence band has a characteristic double-dome shape (with a dome on top of another) and a maximum in $K$. 
In the conduction band, the two bottom bands almost coincide in the $M-K$ path and present two minima close to or at $\Gamma$.
We verified that the bottom conduction in the $\delta = 2$ family does fall in the $\Gamma-M$ high symmetry line (see Appendix J).

\section{Conclusions}

To conclude, we have demonstrated that in hBN bilayers there are five stackings that are invariant under rotations of 120$^\circ$ like the pristine hBN monolayers.
We have listed the symmetry groups of these stackings, shown how to construct them and how to transform one into another and we have introduced a physically informative nomenclature allowing to identify them unambiguously.
We also have provided a precise definition of the twist angle ($\theta$ or $\theta^\prime$ depending on the stacking).
All this contrasts with graphene bilayers, where only two stackings can be defined.
Our nomenclature is completely general and can be applied to any homobilayer formed of hexagonal 2D materials (twisted as well as untwisted).
Even though corrugation and domain relaxation have to be expected in experimental realization of these systems~\cite{Guinea2019,Woods2021,Walet2021, Moore2021},  
 these structural modifications will still be constrained by the stacking sequence. 
By performing DFT simulations, we have done a thorough study of the electronic structure of hBN bilayers taking into account both its dependence on the stacking sequence and the twist angle. 
In the first case, we have traced a correlation between the atom-on-atom coincidences and some characteristics of the states which form the gap. 
In the second case, we have shown that the gapwidth is always indirect irrespective of the twist angle and it decreases for decreasing $\theta$ or for increasing $\theta'$, differently from what previously predicted on the basis of less sophisticated simulation schemes~\cite{Lee2021}.
Finally we have identified the structural parameter  $\delta=|p-q|$ which allows to classify bilayers into families with similar band structures.
The stacking- and angle-dependent properties discussed in this letter have special importance in possible twistronic applications.
In fact these mechanisms are expected to have a strong impact on the optical properties of these bilayers and in particular on the direct manipulation of interlayer excitons which can be stabilized through the application of an external field. 

\begin{acknowledgments}
The authors are thankful to Dr. F. Paleari for fruitful discussions and the dedicated analysis tools he provided.
They also acknowledge the contribution of F. Ducastelle who seeded this work.
Finally, they acknowledge funding from the European Union’s Horizon 2020 research and innovation program under grand agreement N$^\circ$ 881603 (Graphene Flagship core 3) and 
from the French National Agency for Research (ANR) under the projects EXCIPLINT (Grant No. ANR-21-CE09-0016).
\end{acknowledgments}

\section{Appendices}

\subsection{A: Asymetric honeycomb supercells}

As presented in the main article, 
we choose the two primitive vectors of the boron nitride monolayer ${\bf a}_1$
and ${\bf a}_2$ forming an angle of 60$^\circ$
and define the three vectors separating the nitrogen and the boron
sublattices like:
\[
\vectau_1 = +\veca_1/3+\veca_2/3 \quad\quad
\vectau_2 = \vectau_1-\veca_1 \quad\quad
\vectau_3 = \vectau_1-\veca_2
\]
A boron atom is located at the origin of the honeycomb
and nitrogen is located at
$\vectau_1$.
A new periodic super-lattice is constructed with the
new translational vectors
$\vecA_1$ and $\vecA_2$ written on the
basis $\lbrace\veca_1,\veca_2\rbrace$ like
\begin{equation}
  \label{sm:equ:vectorA}
  \mathbf{A}_i=\sum_j M_{ij}\mathbf{a}_j.
\end{equation}
In the bilayer system, the hexagonal supercell for the lower layer has been arbitrarily
chosen as the one produced by the matrix 
\begin{equation}
  \label{sm:equ:rightlower}  
  \text{M}^{(q,p)}
  = 
  \begin{bmatrix}
    q & p  \\
    -p & p+q 
  \end{bmatrix}
\end{equation}
and the upper layer is developed either with
\begin{equation}
  \label{sm:equ:rightupper}  
  \text{M}^{(p,q)}
  = 
  \begin{bmatrix}
    p & q  \\
    -q & p+q 
  \end{bmatrix}
\end{equation}
or with
\begin{equation}
  \label{sm:equ:rightupperprime}  
  \text{M}^{(-q,p+q)}
  = 
  \begin{bmatrix}
    -q & p+q  \\
    -p-q & p 
  \end{bmatrix}\,.
\end{equation}
In all these cases, $p$ and $q$ are integers.
The vertical mirror
planes along the $[ 1\, 1]$ and $[1\,0]$ 
directions of the supercell are lost only if
\[
p \neq 0,\, q\neq 0 \text{ and } p\neq q
\]
then, we call such supercell \emph{asymmetric}.
These are the supercells considered in this work because they lead to twisted bilayers.

Lastly, the $\{p,q\}$ integers define also the parameter
length, the surface $\Omega$ and the numer of atoms $\mathrm{N_{at}}$  of the three supercells
\begin{align}
  \lvert\vecA_i\rvert&= a\sqrt{p^2+q^2+pq} \\
  \Omega &= \Omega_0\left( p^2+q^2+pq \right) \\
  \mathrm{N_{at}}&=2\left( p^2+q^2+pq \right) 
\end{align}
where \(\Omega_0=\frac{a^2\sqrt{3}}{2}\) is the surface,
and $a$ is the cell
parameter of the honeycomb primitive cell.

As we mention in the main article, the origin of a generic $(k,s)$ supercell
can be set either on an atom or on the center of a hexagon of the
underlying honeycomb lattice.
We want to analyze what happens at the direct-space high-symmetry points
\((0\,0)\),
\((\frac{1}{3}\,\frac{1}{3})\) and
\((\frac{2}{3}\,\frac{2}{3})\)
of the supercell where the axes of order-3 rotation symmetry pass (cfr. below).
These points are highlighted with red dots in
Figure~1 of the main article.
Using (\ref{sm:equ:rightupper})
we write
\begin{align}
  \left(\frac{X}{3} \, \frac{X}{3} \right) &=
  \frac{X}{3}\vecA_1+\frac{X}{3}\vecA_2  \\ 
  &= \frac{X}{3}\left(k-s\right)\veca_1+
  \frac{X}{3}\left(k+2s\right)\veca_2
  \label{sm:equ:hspoint}
\end{align}
where the integer $X=1$ or $2$ selects the supercell high symmetry point. 
Let us introduce now the integer parameter $\alpha$ defined as
\[ k-s=3t+\alpha\] 
with $t \in \mathbb{Z}$, so only~$-1$,~$0$ and~$1$ are meaningful values of $\alpha$.
Using it in equation (\ref{sm:equ:hspoint}), we get
\begin{align}
    \left(\frac{X}{3} \, \frac{X}{3} \right)
  &=\frac{X}{3}\left(3t+\alpha\right)\veca_1+
  \frac{X}{3}\left(3t+3s+\alpha\right)\veca_2 \\
  &= \underbrace{ Xt\,\veca_1+X(t+s)\veca_2 }_{=\mathbf{R}}
  +\frac{X\alpha}{3}\left(\veca_1+\veca_2\right)
\end{align}
where \(\mathbf{R}\) is a honeycomb lattice vector.
Therefore, if $\alpha=-1$ and $X=1$, the site located in
\((\frac{1}{3} \, \frac{1}{3} )\) of the \emph{supercell} will
coincide with the site located at  $( -\frac{1}{3} \, -\frac{1}{3} ) = (\frac{2}{3} \, \frac{2}{3} )$
of the \emph{primitive cell} of the honeycomb lattice, and vice-versa if $X=2$.
But if
$\alpha=+1$, the site in 
\((\frac{1}{3} \, \frac{1}{3} )\) will coincide 
with the site in \((\frac{1}{3} \, \frac{1}{3} )\)
of the primitive cell, and the same for $X=2$.
Actually, we demonstrate below in the Supplementary Materials
that the case \(\alpha=0\) is irrelevant.

Lastly, it is easy to demonstrate that if
a given supercell \( (p,q)\) has a \(\alpha=+1\) parameter, then
the supercells
\((q,p)\)
and 
\((-q,p+q)\) have a \(\alpha=-1\) parameter
(and inversely).

\subsection{B: Stacking geometries}

\begin{table*}
  \centering
  \renewcommand{\arraystretch}{1.3}
  \begin{tabular}{r | c @{\hspace{5mm}} ccc | c @{\hspace{5mm}} ccc | l }
   supercell & $\alpha$ & $(0 \,0)$ & $(\frac{1}{3}\,\frac{1}{3})$ & $(\frac{2}{3}\,\frac{2}{3})$ & $\alpha$ &     $(0\,0)$ & $(\frac{1}{3}\,\frac{1}{3})$ & $(\frac{2}{3}\,\frac{2}{3})$ & name of the bilayer obtained \\
   \cline{1-9}
   $(q,p)_\text{B}$ & -1 & B & H & N & +1 & B & N & H & by stacking on the $(q,p)_\text{B}$ \\
   \hline
   $(p,q)_\text{B}$ &    & B & N & H &      & B & H & N & BB$(q,p)$ \\
   $(p,q)_\text{N}$ &+1& N & H & B & -1 & N & B & H & BNNB$(q,p)$ \\
   $(p,q)_\text{H}$ &    & H & B & N &     & H & N & B & NN$(q,p)$ \\
   \hline
   $(-q,p+q)_\text{B}$ &    & B & H & N &      & B & N & H & BBNN$(q,p)$ \\
   $(-q,p+q)_\text{N}$ & -1& N & B & H & +1 & N & H & B & BN$(q,p)$ \\
   $(-q,p+q)_\text{H}$ &    & H & N & B &      & H & B & N & BN$(q,p)$ \\
  \end{tabular}
  \renewcommand{\arraystretch}{1}
  \caption{\label{sm:tab:coincidence2}
    Determination of the kind of the sublattices located
    at the high symmetry points used in our construction of bilayers for a generic $\{p,q\}$ pair, and the name of the resulting bilayer.  }
\end{table*}

As we mentioned in the main article, our construction of the moir\'{e} geometries 
requires two integers $\{p,q\}$ and follows the rules:
(i) the lower layer is always defined by the $(q,p)_{\mathrm{B}}$
supercell (origin at boron) and (ii) the upper layer is either defined by the
$(p,q)_{\mathrm{X}}$ cell or the $(-q,p+q)_{\mathrm{X}}$ cell,
where X labels the origin of the
supercell (B = boron, N = nitrogen, H = hexagon center).
As shown in the previous section, the $(p,q)$-on-$(q,p)$ constructions
will always be made of supercells with opposite $\alpha$ parameters, whereas
the $(-q,p+q)$-on-$(q,p)$ constructions will always result from 
supercells with the same
$\alpha$. The Table~\ref{sm:tab:coincidence2} lists the kind of
sublattice (boron, nitrogen atom, or hexagon center) that occurs at
the high symmetry points for both values of $\alpha$ of the lower layer $(q,p)_\text{B}$.

For any choice of $p$ and $q$,
the six possible stackings are:
\begin{enumerate}
\item The $(p,q)_{\mathrm{B}}$-on-$(q,p)_{\mathrm{B}}$
  is a single coincidence structure,
  with B on B at the origin,
  N on hexagon at one of the two high-symmetry points
  and a hexagon on N at the other one.
  There is no hexagon-on-hexagon vertical alignment
  for the single coincidence structures.
  We call this structure the BB$(q,p)$ bilayer.
 \item The $(p,q)_{\mathrm{N}}$-on-$(q,p)_{\mathrm{B}}$
  is a double coincidence structure,
  with N on B at the origin,
  B on N at one of the two high-symmetry points
  and an hexagon-on-hexagon at the other one.
  We call it the BNNB$(q,p)$ bilayer.
\item The $(p,q)_{\mathrm{H}}$-on-$(q,p)_{\mathrm{B}}$
  is again a single coincidence structure,
  with a hexagon on B at the origin,
  B on hexagon at one of the two high-symmetry points
  and an N on N at the other one.
  We call it the NN$(q,p)$ bilayer.  
\item The $(-q,p+q)_{\mathrm{B}}$-on-$(q,p)_{\mathrm{B}}$
  is another double coincidence structure,
  with B on B at the origin,
  N on N at one of the two high-symmetry points
  and an hexagon-on-hexagon at the other one.
  We call it the BBNN$(q,p)$ bilayer.
\item The $(-q,p+q)_{\mathrm{N}}$-on-$(q,p)_{\mathrm{B}}$
  is a single coincidence structure,
  with N on B at the origin,
  N-on-hexagon at one of the two high-symmetry points
  and an B-on-hexagon at the other one.
  We call it the BN$(q,p)$ bilayer.
\item The $(-q,p+q)_{\mathrm{H}}$-on-$(q,p)_{\mathrm{B}}$
  is a single coincidence structure,
  with a hexagon on B at the origin,
  N on hexagon at one of the two high-symmetry points
  and an B on N at the other one.
  It is the same geometry than the BN$(q,p)$ above.
\end{enumerate}
Finally, since the stacking 6 leads actually to the same structure as 
stacking 5, for each $\{p,q\}$ pair of integer we construct five and only five different
structures that preserve the atom-on-atom vertical alignments.

\subsection{C: Moir\'{e} stacking angles}

\begin{figure}[b]
  \centering
  \includegraphics[width=0.95\linewidth]{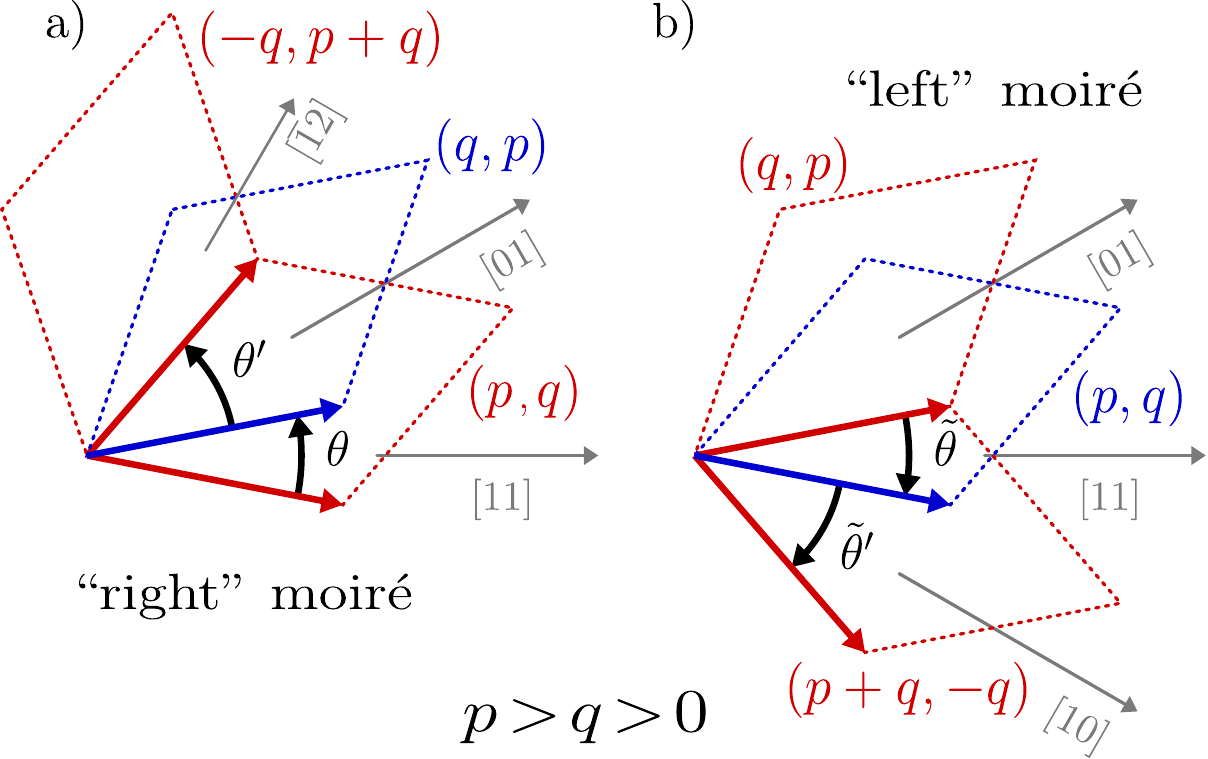}
  \caption{
    \label{sm:fig:angles}
    a) The angles $\theta$ and $\theta'$,
    based on the $(q,p)$ geometries (that are used in the main article).
    b) The angles
    $\tilde{\theta}$ and $\tilde{\theta}'$ corresponding to the mirror
    images of the previous ones. They are based on the $(p,q)$ geometries. 
  }
\end{figure}

The easiest way to derive the twist angle between two bilayers is by representing the vectors
of the honeycomb lattice with discrete complex numbers.
Here, we adopt the notation~\cite{Mele2010,Shallcross2010}
\(\mathcal{Z}(m,n)  = m z_1+n z_2\) with
\(z_1=1\) and \(z_2=\frac{1}{2}+\frac{\sqrt{3}}{2}i\).
The angles are just the arguments calculated like
\begin{align}
  \exp(i\theta)&=\frac{\mathcal{Z}(q,p)}{\mathcal{Z}(p,q)}
  \\
  \exp(i\theta')&=\frac{\mathcal{Z}(-q,p+q)}{\mathcal{Z}(q,p)}
\end{align}
and depend only on the $\{p,q\}$ pair of integers. This leads to
\begin{align}
  \tan \theta_{ \{p,q\} }  &=\sqrt{3} \frac{p^2-q^2}{p^2+q^2+4pq}  \label{sm:equ:rightangles}   \\
  \tan \theta'_{ \{p,q\} } &=\sqrt{3} \frac{q^2+2pq}{2p^2-q^2+2pq}\label{sm:equ:rightangles2}
\end{align}
which are given in the main article.
Since the $p$ and $q$ indices can take any integer value,
the angles are always defined modulo 60$^\circ$.
The constructed supercells and the
resulting angles $\theta$ and
$\theta'$ are drawn in Figure~\ref{sm:fig:angles}.a.

So far, the vectors defined by (\ref{sm:equ:vectorA})
have been developed 
on the \(\lbrace \veca_1,\veca_2\rbrace\) honeycomb lattice basis, but
we could have chosen either to develop them on the 
\(\lbrace \veca_2-\veca_1,-\veca_1\rbrace\) basis and then work
with the $\{-p-q,p\}$ pair,
or on the
\(\lbrace -\veca_2,\veca_1-\veca_2\rbrace\)
basis, and work with the
$\{q,-p-q\}$ pair.
So, definitions~\eqref{sm:equ:rightangles} and~\eqref{sm:equ:rightangles2}
are not unique and the angles could have also been defined as
\begin{align}
  \tan \theta_{ \{-p-q,p\} }  &=\sqrt{3} \frac{q^2+2pq}{-2p^2+q^2-2pq}  \label{sm:equ:rightanglesalter}  \\
  \tan \theta'_{ \{-p-q,p\} } &=\sqrt{3} \frac{-p^2-2pq}{-p^2+2q^2+2pq}  \label{sm:equ:rightanglesalter2}
\end{align}
or
\begin{align}
  \tan \theta_{ \{q,-p-q\} }  &=\sqrt{3} \frac{p^2+2pq}{-p^2+2q^2+2pq}  \label{sm:equ:rightanglesalter3} \\
  \tan \theta'_{ \{q,-p-q\} } &=\sqrt{3} \frac{-p^2+q^2}{p^2+q^2+4pq}  \label{sm:equ:rightanglesalter4}
\end{align}
which are also valid formulations. It is trivial to show that
for any $\theta$ of equations~\eqref{sm:equ:rightangles}, \eqref{sm:equ:rightanglesalter}, or \eqref{sm:equ:rightanglesalter3} and for
any $\theta'$ of equations~\eqref{sm:equ:rightangles2}, \eqref{sm:equ:rightanglesalter2}, or \eqref{sm:equ:rightanglesalter4}, 
the following equality 
\[
 \theta'  = -\theta +  \frac{n \pi}{3}
\]
holds for an integer $ n\in \mathbb{Z}$.
In order to avoid confusion and give a
non ambiguous definitions of our moir\'{e} structures,
we decide arbitrarily
to adopt definitions~\eqref{sm:equ:rightangles} and~\eqref{sm:equ:rightangles2},
and to impose 
\[p>q>0.\]
In this situation, the vectors $p\veca_1+q\veca_2$ and
$q\veca_1+p\veca_2$ lie in the
\(\lbrace \veca_1,\veca_2\rbrace\) angular sector, and the vector
$-q\veca_1+(p+q)\veca_2$ lie in the
\(\lbrace \veca_2,\veca_2-\veca_1\rbrace\)
angular sector. As a consequence
\[
\theta,\theta'\in
{\left]0,\frac{\pi}{3}\right[}
\text{ \quad and\quad } \theta+\theta'=\frac{\pi}{3}
\]
implying that BB$(q,p)$, BN$(q,p)$ et NN$(q,p)$ have an angle $+\theta>0$
and BBNN$(q,p)$, BNNB$(q,p)$
have an  angle $-\theta'<0$. These five stackings are chiral structures, that
we decide to name 
``right-hand'' moir\'{e} bilayers.

To construct the enantiomers of the ``right-hand'' moir\'{e} bilayers above, we
have to transform the vectors $\vecA_1$ defining the
hexagonal supercells (\ref{sm:equ:vectorA}).
They are mirrored respect the \([1\,1]\)
crystallographic direction of the primitive honeycomb lattice cell, as shown in the
Figure~\ref{sm:fig:angles}.b.
The lower layer of a ``left'' moir\'{e} is now carried by the supercell $\text{M}^{(p,q)} $
and the upper layer is developed either on the $\text{M}^{(q,p)} $ 
or the $\text{M}^{(p+q,-q)}$ one, still within the constraint $p>q>0$.
The corresponding twist angles are now
\begin{align}
  \exp(i\tilde{\theta})&=\frac{\mathcal{Z}(p,q)}{\mathcal{Z}(q,p)}
  \\
  \exp(i\tilde{\theta}')&=\frac{\mathcal{Z}(p+q,-q)}{\mathcal{Z}(p,q)}
\end{align}
leading to $ \tilde{\theta}  = -\theta $ and $\tilde{\theta}' = -\theta' $
then
\[
\tilde{\theta},\tilde{\theta'}\in
{\left]-\frac{\pi}{3},0\right[}
\text{\; and \;\;}
\tilde{\theta}+\tilde{\theta}'=-\frac{\pi}{3}.
\]
As a result, the ``left'' BB$(p,q)$, BN$(p,q)$ and NN$(p,q)$ have an angle 
$-\theta<0$, and the ``left" BBNN$(p,q)$ and BNNB$(p,q)$ 
have an angle $+\theta'>0$.

In absence of any magnetic field, the ``right-hand'' and ``left-hand''
corresponding stackings exhibit exactly the same electronic
properties. That is why we restricted our study to the ``right-hand'' ones.

\subsection{D: Redundancy of the case $(p-q=3t)$}

\begin{figure}
  \centering
  \includegraphics[width=0.49\textwidth]{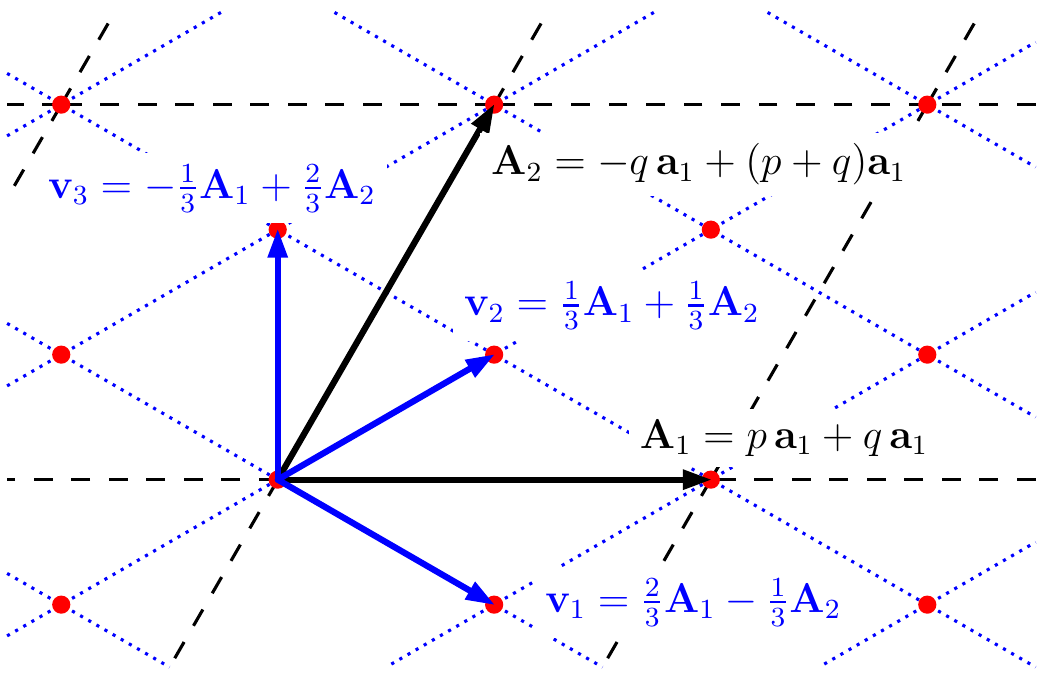}
  \caption{
    \label{sm:fig:alphazero}
    The upper layer asymmetric supercell
    \((p,q)\) with \( p=q+3t\). 
    It is always possible
    to construct a smaller supercell since
    ${\bf v}_1$, ${\bf v}_2$ and ${\bf v}_3$ are vectors of
    the honeycomb lattice.
    In other words, the twisted bilayer geometries constructed from the $(q,q+3t)$ 
    supercell are not primitive cells of the moir\'{e}.
  }
\end{figure}

\begin{figure}[b]
  \includegraphics[width=0.99\linewidth]{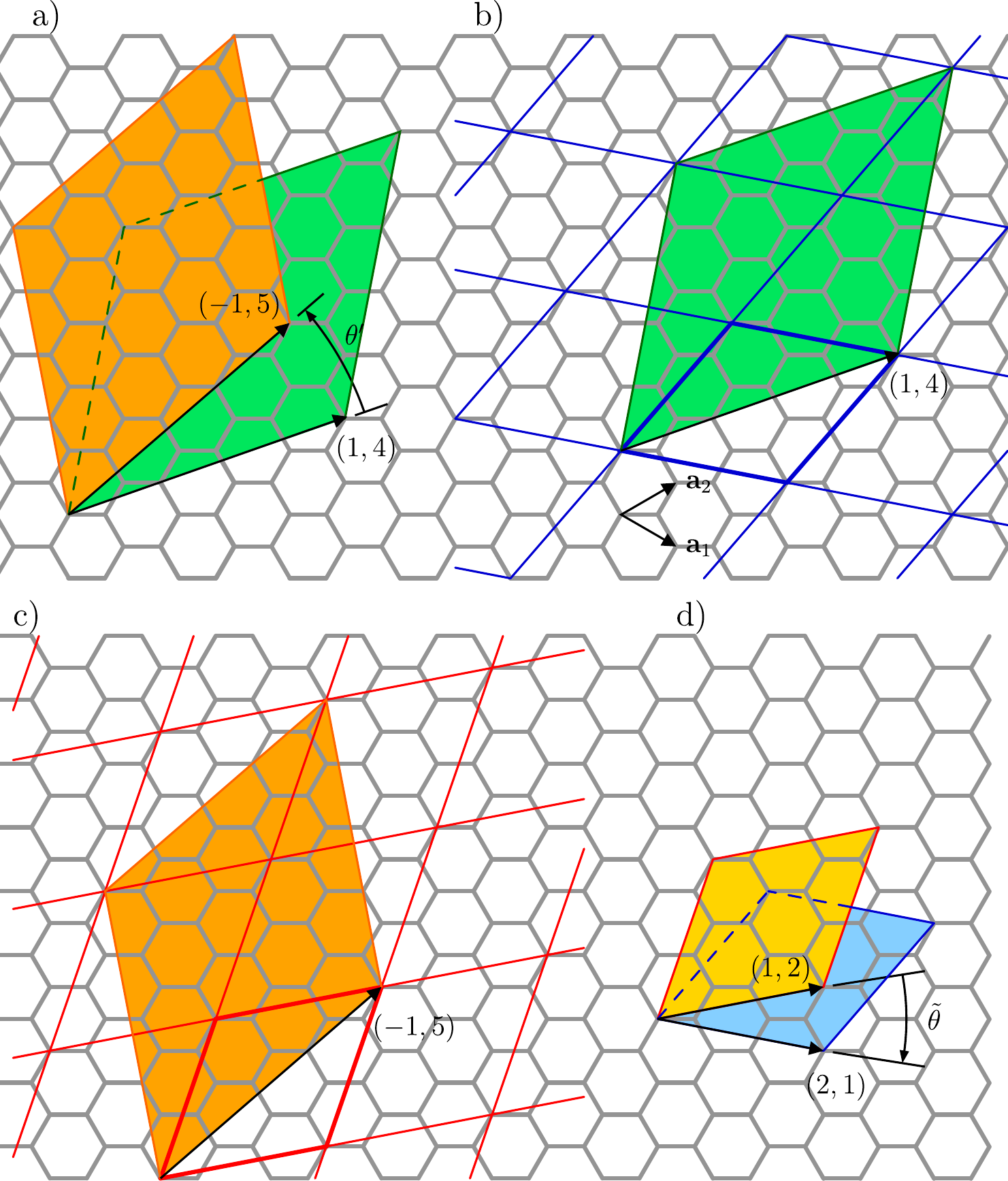}
  \caption{
    \label{sm:fig:useless_example}
    a) Construction of the moir\'{e} bilayer based on the $(1,4)$
    supercell for the lower layer and the $(-1,5)$ for the upper layer.
    b) The lower supercell can be tessellated by the $(2,1)$ smaller supercell.
    c) The upper one is also a tessellation of the $(1,2)$ supercell.
    d) The angle  of the ``left-hand'' small moir\'{e} is the same as
    that of the large non-primitive moir\'{e}
    $\tilde{\theta}_{ \{2,1\} }=-\theta'_{ \{1,4\} }$.
  }
\end{figure}

The case \(\alpha=0\)  corresponds to moir\'{e} $(p,q)$ supercells
where $p-q=3t$ and $t$ is an integer. So
\begin{equation}
  \label{sm:equ:uselesscell}  
  \begin{bmatrix}
    q+3t & q 
    \\
    -q & 2q+3t
  \end{bmatrix}
  =(q+3t,q)\text{ supercell.}
\end{equation}

As we sketched in Figure \ref{sm:fig:alphazero},
starting from the 
vectors \(\vecA_1\) and \(\vecA_2\), we can define new shorter vectors
\begin{align}
  \label{sm:equ:v1v2v3}
  \vecv_1 &= \frac{2}{3}\vecA_1-\frac{1}{3}\vecA_2
  =(q+2t)\,\veca_1-t\,\veca_2 \\
  \vecv_2 &= \frac{1}{3}\vecA_1+\frac{1}{3}\vecA_2
  =t\,\veca_1+(q+t)\,\veca_2 \\
  \vecv_3 &= -\frac{1}{3}\vecA_1+\frac{2}{3}\vecA_2
  =(-q-t)\,\veca_1+(q+2t)\,\veca_2 
\end{align}
and since $q$ and $t$ are integers, the vectors $\vecv_i$ are
honeycomb bravais lattice vectors.
In this situation, the supercell defined by the indices of the
vector $\vecv_3$ (for example) is
\begin{equation}
  \label{sm:equ:smallercell}  
  \begin{bmatrix}
    -q-t &  q+2t
    \\
    -q-2t & t
  \end{bmatrix}
  =(-q-t,q+2t)\text{ supercell}
\end{equation}
which is also an asymetric hexagonal supercell, three times smaller than
the original $(q+3t,q)$ one.

Moreover, the twist angles (\ref{sm:equ:rightangles}) calculated with
$p$ and $q$ indices (when $p=q+3t$) are
\begin{equation}
\begin{split}
  \tan\theta_{ \{q+3t,q\} }&=\sqrt{3}
  \frac{3t^2+2qt}{2q^2+3t^2+6qt}
  \\
  \tan\theta^\prime_{ \{q+3t,q\} }&=\sqrt{3}
  \frac{q^2+2qt}{q^2+6t^2+6qt} \nonumber
\end{split}\end{equation} 
and it is staightforward to verify than these two tangents
are exactly the same if we calculate them with the $-q-t$ and $q+2t$ indices.

To summarize, (i) the $\{q+3t,q\}$ set leads to non primitive
moir\'e supercells,
and (ii) it is always possible to use the $\{-q-t,q+2t\}$ pair
which gives the same twist
angles but in three times smaller supercells.
As an illustration of it, in Figure~\ref{sm:fig:useless_example} we have drawn
the example of the construction of the $(-1,5)$-on-$(1,4)$ moir\'{e} and its
reduction to the $(1,2)$-on-$(2,1)$ ``left-hand'' moir\'{e} bilayer.


\begin{figure*}
  \includegraphics[width=0.90\textwidth]{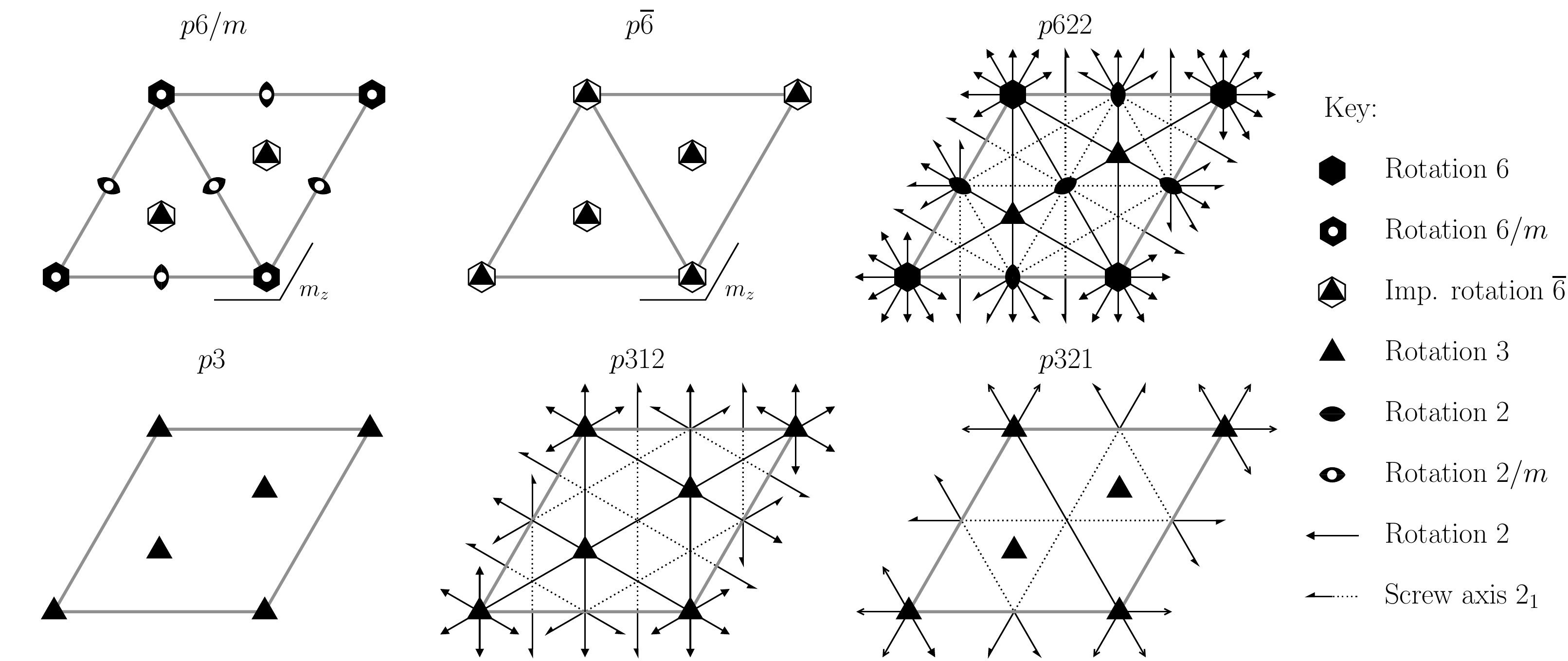}
  \caption{
    The
    graphene and hBN
    moir\'{e} bilayers belong to one of these layer groups (adapted from \cite{Kopsky:tableE}).
    The trivial group $p1$ is not shown.
  }
\label{sm:fig:layergroups}
\end{figure*}

\subsection{E: Completeness of the description with $(q,p)$ pairs}
In this section we show that all possible hexagonal supercells can be expressed by an appropriate choice of the $(q,p)$ pair of integers.

Let us first highlight the relation between the $(q,p)$ pair and the translational symmetry of the bilayer.
The bottom layer is generated by the two vectors $\mathbf{a}_1$ and $\mathbf{a}_2$ spanning an angle of $60^\circ$.
On top of it, another hexagonal layer is rotated by some angle around a pivotal point which is placed at the origin of the axes for sake of simplicity. 
Its primitive vectors are $\mathbf{c}_1 = \mathcal{R}\mathbf{a}_1$ and $\mathbf{c}_2 = \mathcal{R}\mathbf{a}_2$ defined through the rotation matrix $\mathcal{R}$.
If the resulting structure has translational symmetry, there must be replicas of the pivotal point and they must form a supercrystall generated by a unitary supercell with hexagonal symmetry.
By definition, it must be possible to express the unitary supercell as an integer linear combination of the unitary vectors of each layer.
Let it be $\mathbf{A}= q \mathbf{a}_1 + p \mathbf{a}_2$ if expressed in the lower layer unitary vectors, then it coincides either to $\mathbf{C} = p \mathbf{c}_1 + q \mathbf{c}_2$ or to $\mathbf{C} = -q \mathbf{c}_1 +(p+q) \mathbf{c}_2$ in the the upper layer, depending on weather it is rotated by $\theta$ or $\theta'$, as explained in the previous appendices. 

In the main text we limited our study to ``right-hand" bilayers for which $p>q>0$, and twist angles are $+\theta$ and $-\theta^\prime$. These two angles are defined inside the open interval $] 0^\circ, 60^\circ[$ and their tangents are comprised in the interval $]0,\sqrt{3}[$.
Let us introduce the real number $\xi \in ]0,1[$.
Formulae~\eqref{sm:equ:rightangles} and~\eqref{sm:equ:rightangles2} can be expressed in terms of this quantity
\begin{eqnarray}
\tan \theta &=& \sqrt{3}\frac{1-\xi^2 -1 }{ \xi^2 + 4\xi +1 } \\ 
\tan \theta' &=& \sqrt{3}\frac{\xi^2 + 2\xi }{-\xi^2 + 2\xi + 2} \,.
\end{eqnarray}
These two expressions are bijective relations mapping $]0,1[$ to $]0,\sqrt{3}[$.
Therefore, one can invert these relations and since the tangent is also invertible, it is possible to establish a bijection $]0^\circ,60^\circ[ \rightarrow ]0,1[$ mapping any $\theta$ or $\theta'$ angle to $\xi$.
If $\xi$ is rational, one can always find a pair of integers $(q,p)$ such that $\xi = q/p$. 
On the other hand, if it is not rational, then it will not be possible to find any pair of integers $(q,p)$ corresponding to the chosen angle, but in this case the system will have no translation symmetry.

\subsection{F: Layer groups of moir\'{e} structures}

In Figure~\ref{sm:fig:layergroups}, we report graphical representations
of the symmetries of the layer group used in this Appendix.
The layer group of a graphene monolayer asymmetric
supercell is the \(p6/m\),
neglecting translations occurring inside the defined cell.
For a boron nitride (or a transition metal dichalcogenide) supercell, the layer group is \(p\overline{6}\)	
\cite{Kopsky:tableE}.
Both groups contain order-3 or order-6 rotations axis along $z$, located
at the high symmetry points of the cell:
\(\left(0\,0\right)\),
\(\left(\frac{1}{3}\,\frac{1}{3}\right)\) and
\(\left(\frac{2}{3}\,\frac{2}{3}\right)\).
When stacking two supercells like described in the previous sections,
these axes are coincident, and 
the rotations are always preserved.
Thus the 2D crystal systems remain hexagonal.

By looking at Table~\ref{sm:tab:coincidence2} and by replacing all occurrences of B and N by C, it is easy to derive all the stackings of graphene bilayers, 
however the result is highly redundant.
Actually, by taking the origin of all the supercells only on the site corresponding to B atoms in hBN, it is possible to sort out identical geometries from the beginning.
In this case, the 
$(-q,p+q)$-on-$(q,p)$ structure geometry
always shows 
one ``hexagon-on-hexagon'' vertical alignment with
an order-6 rotation axis, and two atom-on-atom vertical alignments with
order-3 rotation axes (\emph{double} sublattice coincidence).
The resulting layer group is the hexagonal \(p622\), that also contains
many in-plane order-2 rotations, oriented along
\([1\,0]\) and \([1\,1]\) crystallographic directions
as well as many $2_1$ screw axes.
Note that to comply with the definitions of layer group as defined in Figure~\ref{sm:fig:layergroups},
the supercell must have the ``hexagon-on-hexagon'' axis is located at the origin. 
This means that supercells constructed as we have done in our work must be translated accordingly.
Differently, the case of $(p,q)$-on-$(q,p)$ structure
exhibits
two ``hexagon-on-atom'' alignments
and one ``atom-on-atom'' alignment
(\emph{single} sublattice coincidence) in the points where order-3 rotation axes pass.
If the structure is constructed like proposed above in
this Supplementary Material, this ``atom-on-atom'' coincidence is correctly located at
the origin. It is worth noticing that there are in-plane order-2 rotations axes,
oriented along the \([1\,0]\) crystallographic directions, passing through the origin.
The symmetry group is \(p321\) for this case.

Let now analyze the symmetry of the hBN moir\'{e} bilayers. 
As explained in the previous sections, the three stackings
BB$(q,p)$,
NN$(q,p)$,
and BN$(q,p)$ correspond geometrically to the graphene bilayer with
\emph{single} sublattice coincidence.
Note that, as previously, the NN stacking must be
translated in such a way that
the ``atom-on-atom'' vertical coincidence is placed at the origin, while this is not needed for the other two stackings that result constructed consistently.
The BB and the NN stacking geometries keep the
in-plane order-2 rotations axes along \([1\,0]\). Therefore their layer group is
also the \(p321\). However, in the BN stacking case, the coincident atoms
are now chemically different and the order-2 rotations are lost.
The group is the simplest hexagonal \(p3\).

The last two hBN moir\'{e} stackings are the
BBNN$(q,p)$ and the BNNB$(q,p)$ which correspond geometrically to the graphene
\emph{double}  sublattice coincidence moir\'{e}.
Again, we translate the structures to locate the ``hexagon-on-hexagon''
vertical axis at the origin.
A careful observation of the BNNB$(q,p)$ moir\'{e} geometry
allows us to notice that the in-plane order-2 rotation axes along \([1\,0]\)
and passing through the origin are conserved.
The layer group of the BNNB moir\'{e} stacking is then again the \(p321\).
Differently, in the BBNN$(q,p)$ structure, the in-plane order-2 rotation axes
that are preserved are oriented along the
\([1\,1]\) crystallographic directions. The layer group of symmetry of
BBNN stacking is then the \(p312\).

In this work, we have built structures paying attention to preserve the
vertical atomic coincidence, and consequently the order-3 rotation axes.
However, we can ask ourselves what happens if we stack
a \((p,q)\) or a \((-q,p+q)\)
supercell on a \((q,p)\) cell with a totally random translation between
the layers.
In this scenario, all the
point symmetry operations are lost, and only the translations
are preserved by construction. 
This implies that, although the supercell vectors have the same length and span an angle of 60$^\circ$,
the crystal system is no longer hexagonal.
It is \emph{oblique} and
the layer group is the simplest \(p1\).
In the reciprocal plane, only the \(+\veck / - \veck \) symmetry is conserved,
and consequently
the high-symmetry points K are no longer equivalent.

\subsection{G: Computational details}

\begin{figure}[b]
\centering
\includegraphics[width=0.35\textwidth]{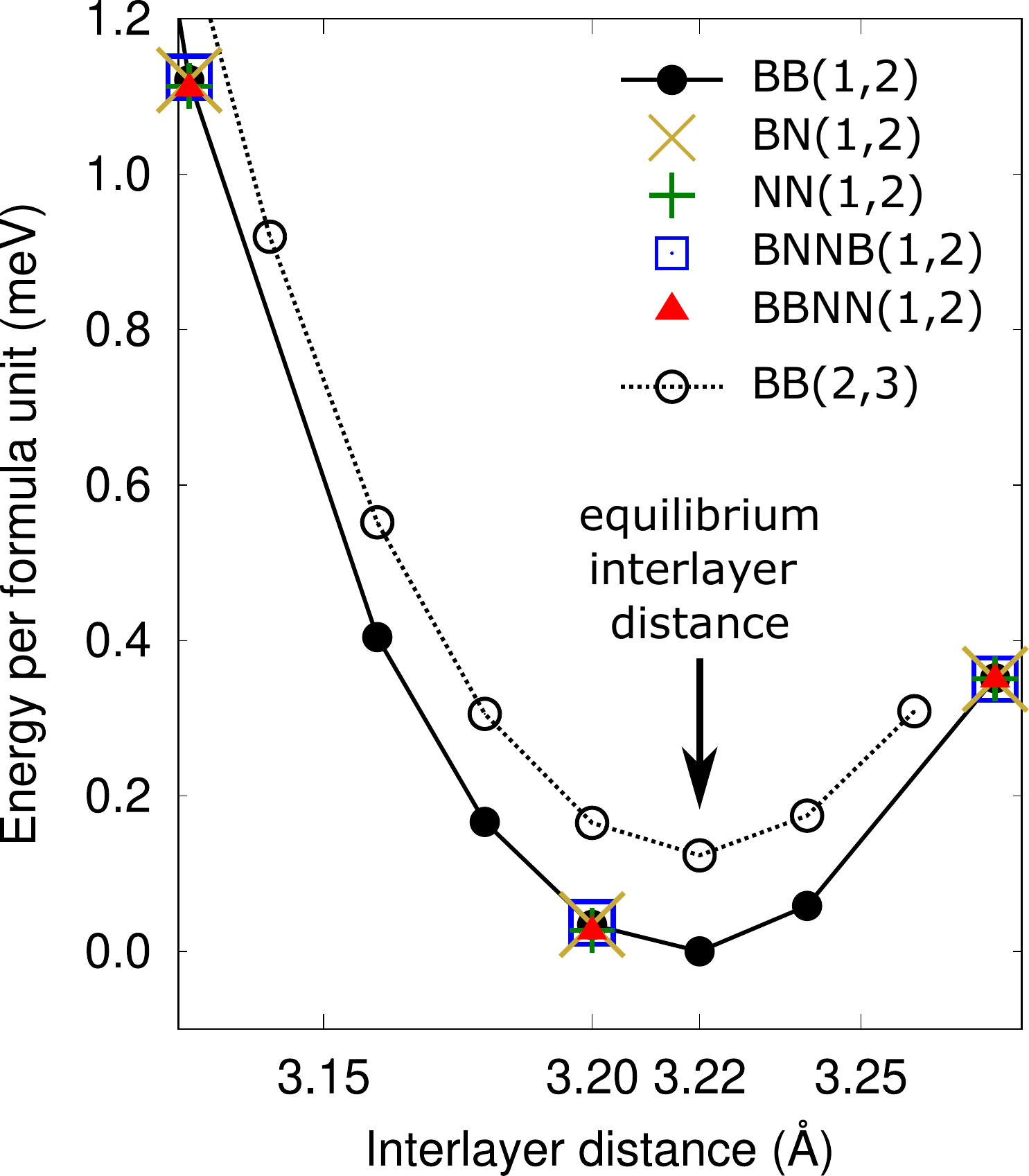}
\caption{Total energy calculation of the five stackings in the (1,2)
  supercell  as a
  function of the interlayer distance $h$. The BB(1,2) is the full black line with black bullets and the BB(2,3) is the dotted line with empty circles. 
  The other (1,2) stackings are superimposed to the BB(1,2) curve almost exactly and are reported with different colors and symbols.} 
\label{fig:interlayer_distances}
\end{figure}

\begin{figure*}
\centering
\includegraphics[width=0.99\textwidth]{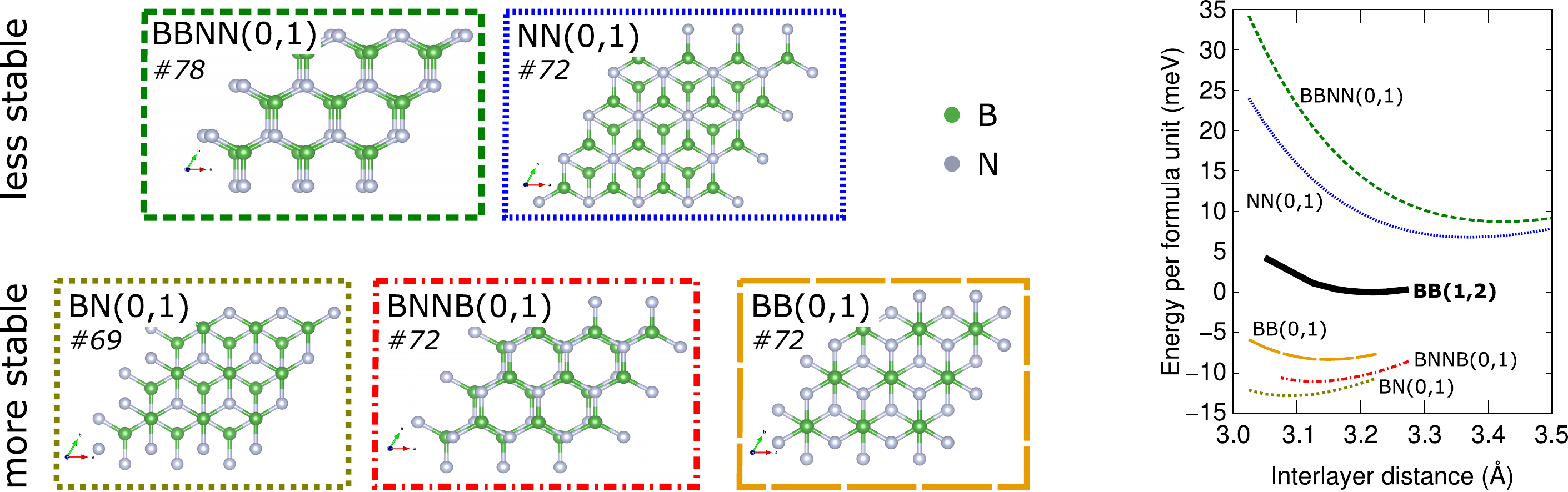}
\caption{The five hexagonal stackings  in untwisted bilayers, their symmetry layer group (\#) and their stability curves with respect to the BB(1,2) twisted bilayer.}
\label{fig:untwisted_structures}
\end{figure*}

Calculations have been done with the free simulation packages Quantum
ESPRESSO~\cite{QuantumEspresso_1, QuantumEspresso_2} (band structure of twisted bilayers) and
ABINIT~\cite{Abinit1,Abinit2} (stability of twisted and untwisted bilayers). 

In both cases norm-conserving pseudopotentials have been used.  We
checked that switching from one software to the other was not
introducing major errors in the main characteristics discussed in the
paper.  In both groups of calculations, the cutoff energy was 30~Ha
and we sampled the Brillouin zone with a Monkorst-Pack grid of
$5\times 5\times 1$ k-points in all supercells ($9\times 9 \times 1$ in the untwisted cases).
We used the PBE exchange correlation potential~\cite{Perdew1996} and included van der Waals corrections via the Grimme-D2 scheme~
\cite{Grimme2006}.
The equilibrium
interlayer distance has been fixed at 3.22~\AA{ }in all bilayers as
detailed below.  The in-plane cell
parameter was $a=2.23$~\AA{ }and no in-plane relaxation has been done.
A cell height $L=$15~\AA{ }has been used in all calculations unless
specified differently. This value has been fixed by paying attention
to the alignment of the $\sigma$ and $\pi$ conduction bands.  In fact,
as already pointed out by several
authors~\cite{Posternak1983,Posternak1984,Blase1994a,Blase1994b,Blase1995,Hu2010,Paleari2019}
the bottom conduction in $\Gamma$ is composed of nearly-free-electron (NFE)
states that extend for several \AA ngstr\"{o}ms above the layer and
thus converge very slowly with the amount of vacuum
 (Appendix I).

To fix the interlayer distance, we calculated the total energy per
unit formula $E(h)$ at different input values of the interlayer distance $h$.
Results are reported in
Figure~\ref{fig:interlayer_distances}. 
 We took the BB(1,2) and the
BB(2,3) bilayers as reference structures. For these bilayers, we sampled $h$ on
a fine grid.  Both bilayers have the energy minimum at $h=3.22$~\AA,
with a negligible energy difference ($\sim$ 0.1 meV per formula unit).
Then we computed $E(h)$ for
the BN(1,2), NN(1,2), BNNB(1,2) and BBNN(1,2) bilayers on a coarser grid and found that the points fell
basically on top of the BB(1,2) curve.  Following this analysis, we
deduced that we can safely fix the equilibrium distance at $h=3.22$~\AA{ }irrespective
of the stacking or the twist angle.
We note however that this value may be inaccurate for very small twist angles that are not investigated in this work.

\subsection{H: Untwisted bilayers}

It is possible to extend the nomenclature we introduced in the main text to untwisted bilayers.
In this case, only the stacking label is meaningful, the $(q,p)$ pair being trivially 1 and 0. 
Note however that at fixed stacking, the symmetry group of the untwisted bilayers (reported in Figure~\ref{fig:untwisted_structures}) differ from that of the twisted ones.
In Figure~\ref{fig:untwisted_structures} we report an image of the structure of the five untwisted stackings and their stability curve $E(h)$ together with that of the
BB(1,2) bilayer.
We observe that the three most stable
untwisted structures, i.e. the BN(0,1), the BNNB(0,1) and the BB(0,1) have a smaller equilibrium
distance, whereas for the two most unstable, the NN(0,1) and the BBNN(0,1), the equilibrium $h$ is
larger, so that the twisted bilayers fall somewhat between the two groups.
This makes sense if one reckons that inside the same twisted
bilayer one can find domains with a local stacking intermediate to the
five untwisted ones.

In experiments it is observed that, far from certain angles, it is
pretty easy to move or twist a BN flake on top of another, and this is
consistent with the negligible energy differences we calculated
between different stackings at fixed angle and between the two
reference calculations with the same stacking sequence.  However when
the twist angle gets close to some specific values, the flake gets
stuck and no further twist is possible.  In fact, the large energy
differences with the untwisted configurations (order of 10 meV per
unit formula) suggest that when approaching small twist angles the
bilayer falls into one of the energetically more favorable
configurations, possibly undergoing large in-plane
deformation to maximize the size of the untwisted
domains.~\cite{Guinea2019, Walet2021, Woods2021, Yao2021a}. 

The equilibrium distances, the total energy per BN pair with respect to the BB(1,2) bilayer and the values of the DFT direct (at K) and indirect band gaps (between valleys close to K and the point M) are reported in Table~\ref{tab:basic_characteristics}.


\begin{table}[b]
\centering
\begin{tabular}{r | cc | cc}
\hline \hline  System & $h$& $E_{\text{BN}}$& $E_{\text{ind}}$& $E_\text{dir}$\\
\hline
 BBNN(0,1)  & 3.425 & 8.7 & 3.957 & 4.037 \\
 NN(0,1)  & 3.375 & 6.8 & 4.345 & 4.037 \\
  BB(2,3) & 3.220 & 0.1 & 4.217 & 4.251 \\
    \textbf{BB(1,2)} & \textbf{3.220} & \textbf{0} & {\bf 4.318} & {\bf 4.394} \\
 BB(0,1)  & 3.150 & -8.3 & 3.950 & 4.436 \\
BNNB(0,1)  & 3.125 & -11.1 & 4.649 & 4.398 \\
BN(0,1) & 3.100 & -12.8 & 4.463 & 4.438 \\
    \hline \hline
\end{tabular}
\caption{Equilibrium interlayer distance $h$ (\AA), total energy per
  formula unit $E_{\text{BN}}$ with respect to the BB(1,2) bilayer (in meV) , smallest indirect gap $E_\text{ind}$ (eV) and energy of the smallest direct transition $E_\text{dir}$ (eV) (direct gap).}
\label{tab:basic_characteristics} 
\end{table}

\subsection{I: Nearly-free-electron states}
\label{sec:appendix_nfe}

\begin{figure}[b]
\centering \includegraphics[width = 0.40\textwidth]{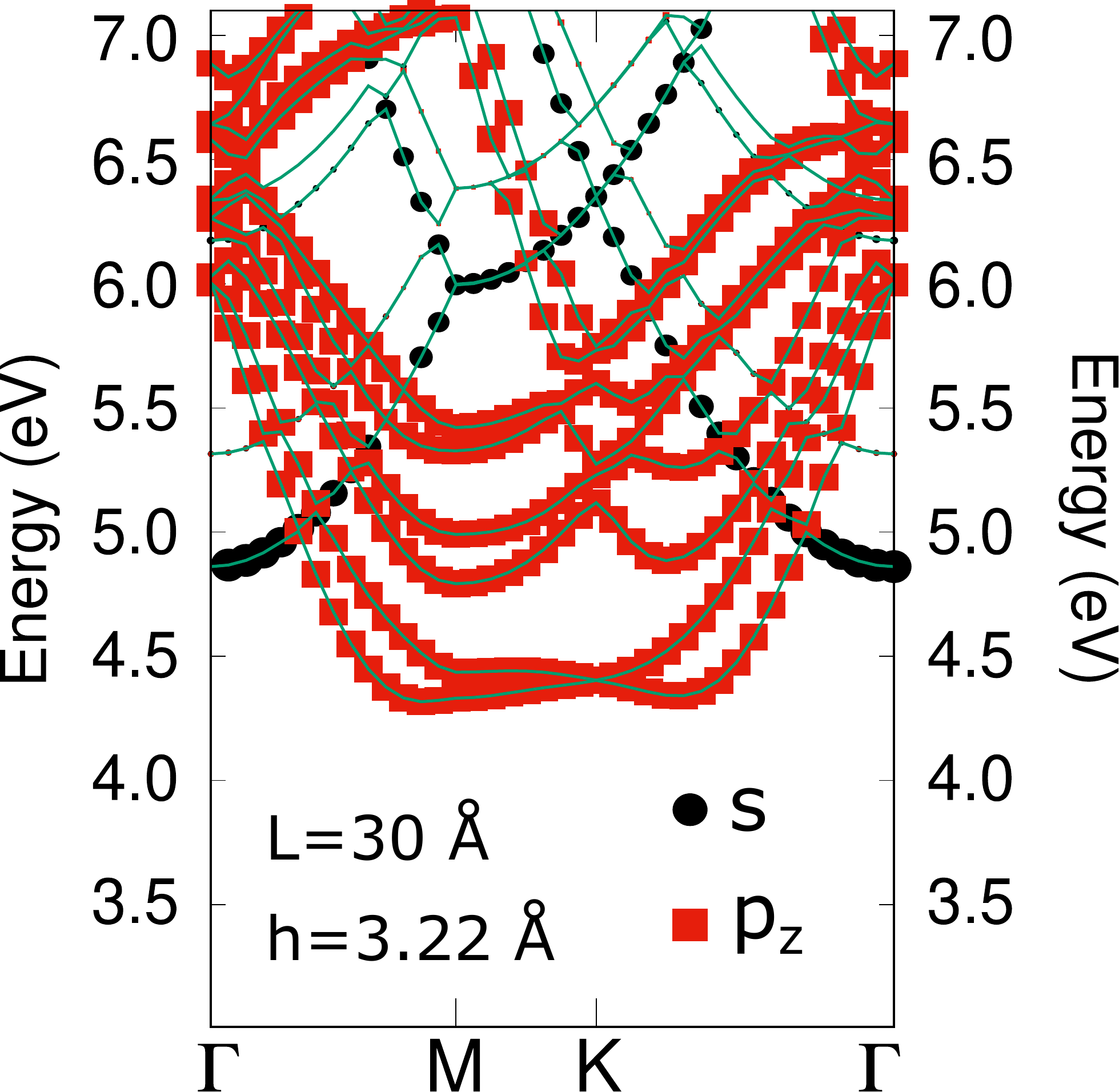}
\caption{Orbital momentum component of the conduction bands of the BN(1,2) bilayer (fat bands).}
\label{fig:fat_bands}
\end{figure}

As already pointed out by Blase and coworkers in the case of bulk
hBN~\cite{Blase1995}, the conduction states at $\Gamma$ converge very
slowly with the amount of vacuum because they correspond to some
unoccupied N-centered nearly-free-electron (NFE) state extending for
several \AA{}ngstr\"{o}ms above the BN layer~\cite{Posternak1983, Posternak1984,Blase1994a,Blase1994b,Blase1995,Hu2010,Paleari2019}.
These NFE states have a neat $3s$ orbital component, as
shown in the fat-band plot reported in Figure~\ref{fig:fat_bands}.

\begin{figure}
\centering \includegraphics[width = 0.49\textwidth]{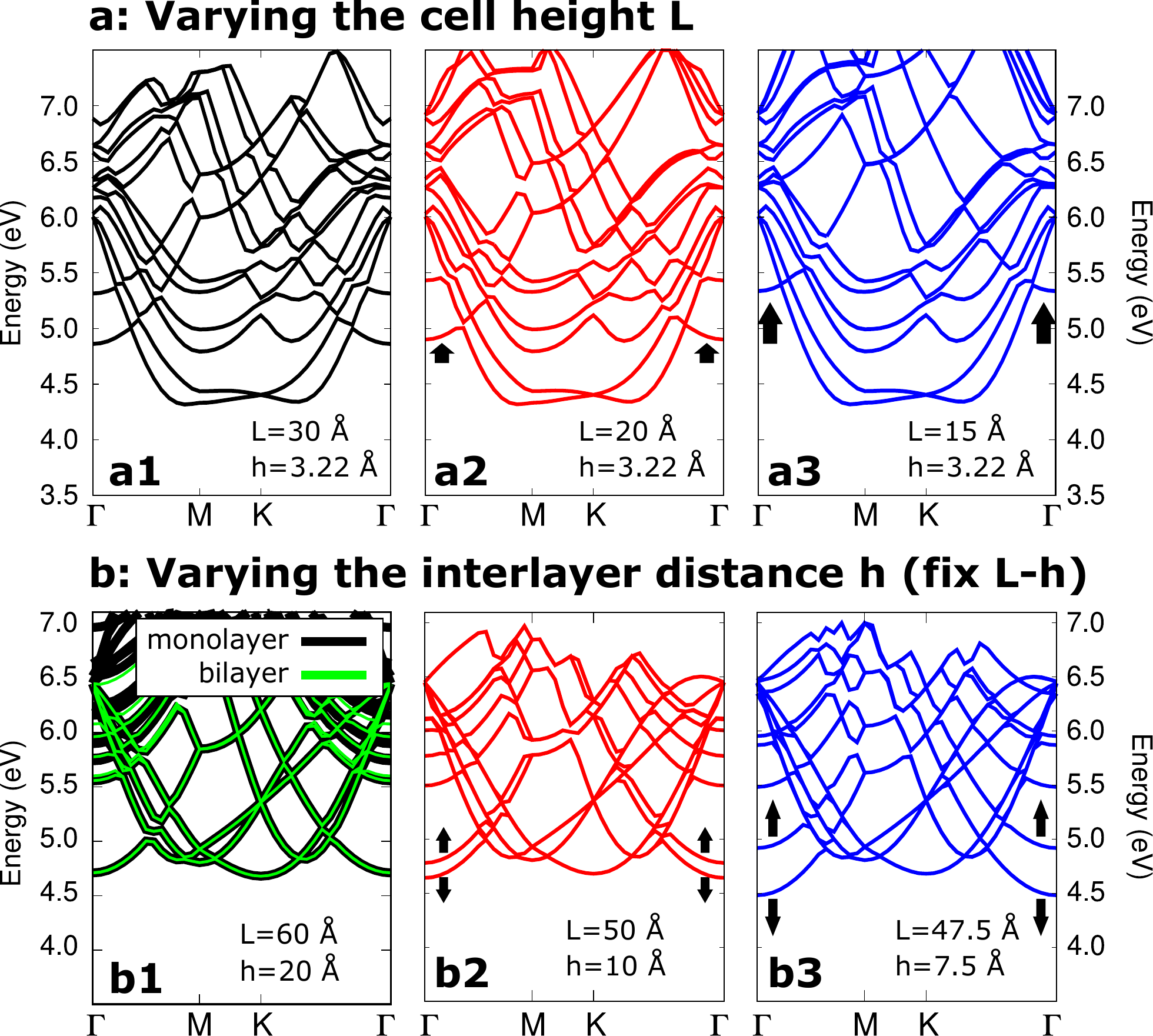}
\caption{The evolution of the NFE states as a function of the simulation parameters in the BN(1,2) bilayer.  \textbf{a}:
  evolution as a function of the cell height $L$ at fixed interlayer dsitance ($h=3.22$~\AA).
  $L = 30$, $20$ and $15$~\AA{ }in panels {\bf a1}, {\bf a2} and {\bf a3} respectively.
     \textbf{b}:
   evolution as a function of the interlayer dsitance $h$ at fixed vacuum ($L-h=40$~\AA).
   $h = 20$, $10$ and $7.5$~\AA{ }in panels {\bf b1}, {\bf b2} and {\bf b3} respectively.
   In panel {\bf b1}, the band structure of the BN(1,2) bilayer (flashy green) is compared with that  of the isolated monolayer (black).}
\label{fig:fel_bands}
\end{figure}

Their alignment with respect to the $\pi$ bands is a delicate issue on
the purpose of this article because the energy difference between the
bottom of the unoccupied $\sigma$ band and the bottom of the
unoccupied $\pi$ band are very close in energy and they may compete in
determining the indirect nature of the gap.  Therefore, it is worth
paying much attention to their convergence.  To this aim, we made a
series of two test calculations in a BN(1,2) bilayer.
First we tested the evolution of these states as a function of the height of the simulation cell at fixed interlayer distance (the three panels of Figure~\ref{fig:fel_bands}a).
This test shows that by reducing the cell height, the NFE states are pushed toward higher energies because of fictitious cell-to-cell interactions. Replicas of the system must be separated of around $L \sim 20$~\AA{  }for the band
dispersion and alignment to be converged.
Note that we decided on purpose to carry out our simulations with a slightly lower value (15~\AA) because the fact of pushing the NFE states to higher energies is not detrimental to our investigation and allows us to reduce the computational workload.

Then we tested the evolution of the NFE states as a function of the interlayer distance leaving a constant amount of vacuum ($L-h$) of 40~\AA, which is largely enough to prevent cell-to-cell interactions.
In the panels of Figure~\ref{fig:fel_bands}.b, we report
three calculations of the BN(1,2) bilayer with a varying interlayer distance (20, 10 and 7.5~\AA{ }respectively in panels b1, b2 and b3).
In the b1 panel, we also plot in black the conduction band of the isolated monolayer in the (1,2) supercell and we verify that it coincides with the $h=20$~\AA{ }bilayer calculation. 
This test demonstrates that moving two layers closer to each other induces a bonding/antibonding splitting of the NFE states which increases as the layers get closer.

Since there is no difference between the interlayer distance separating two layers inside the cell and the space separating replicas of the simulated system, one should pay attention that these two effects (pushing to higher energies and band splitting) happen at the same time.

\subsection{J: Band gap of the $\delta=2$ family}
\begin{figure}
\centering \includegraphics[width=0.49\textwidth]{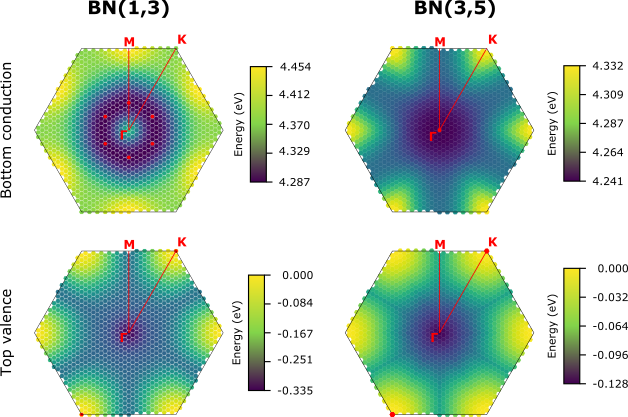}
\caption{Energy surface of the lowest empty band (top panels) and the highest occupied band (bottom panels) of the BN(1,3) and the BN(3,5)  bilayers from left to right.
The top valence and the bottom conduction states are highlighted with red hexagons. } 
\label{fig:band_surfaces}
\end{figure}

\begin{figure}
\centering

\includegraphics[width = 0.49\textwidth]{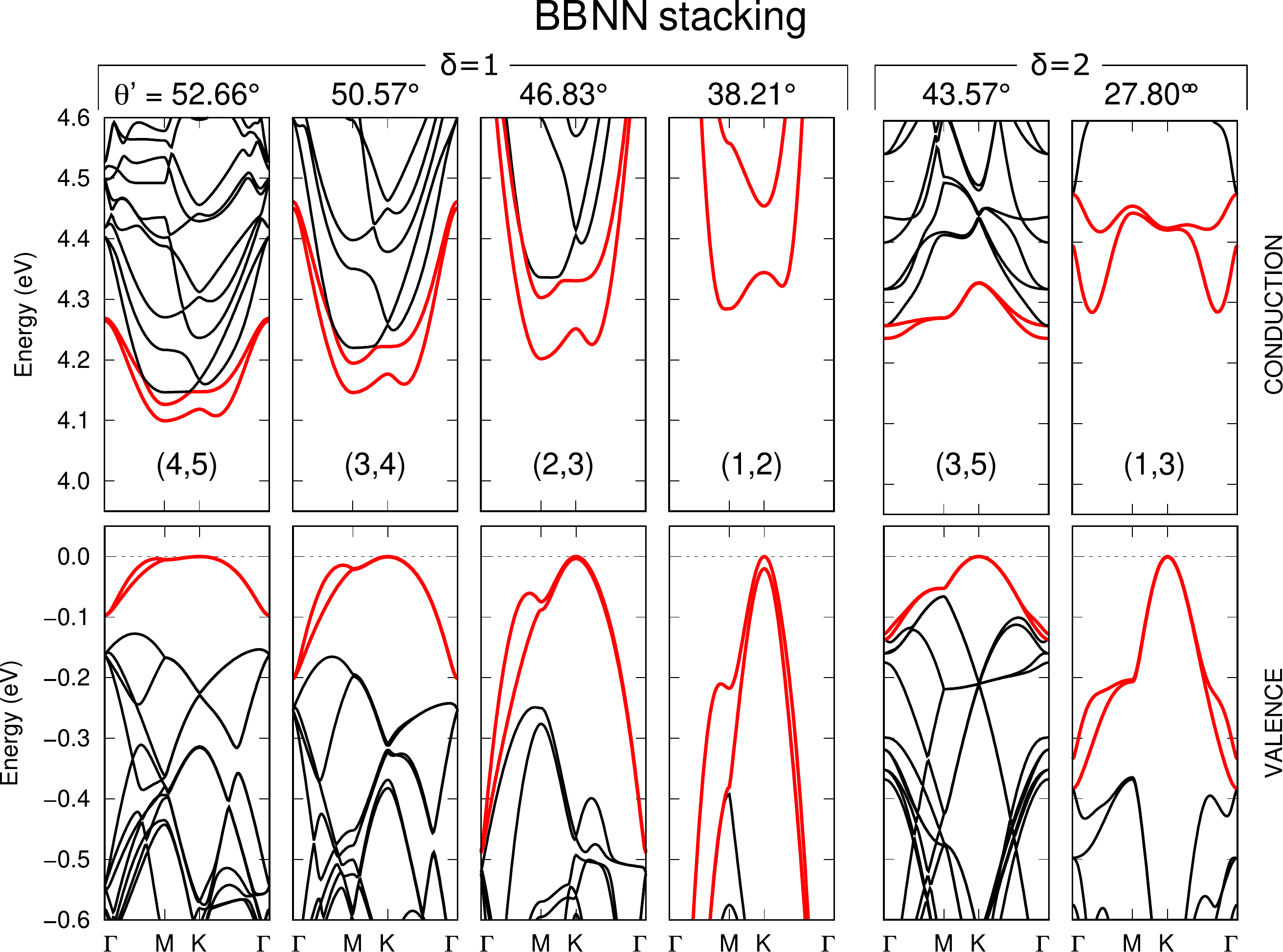}

\vspace{10mm}

\includegraphics[width = 0.49\textwidth]{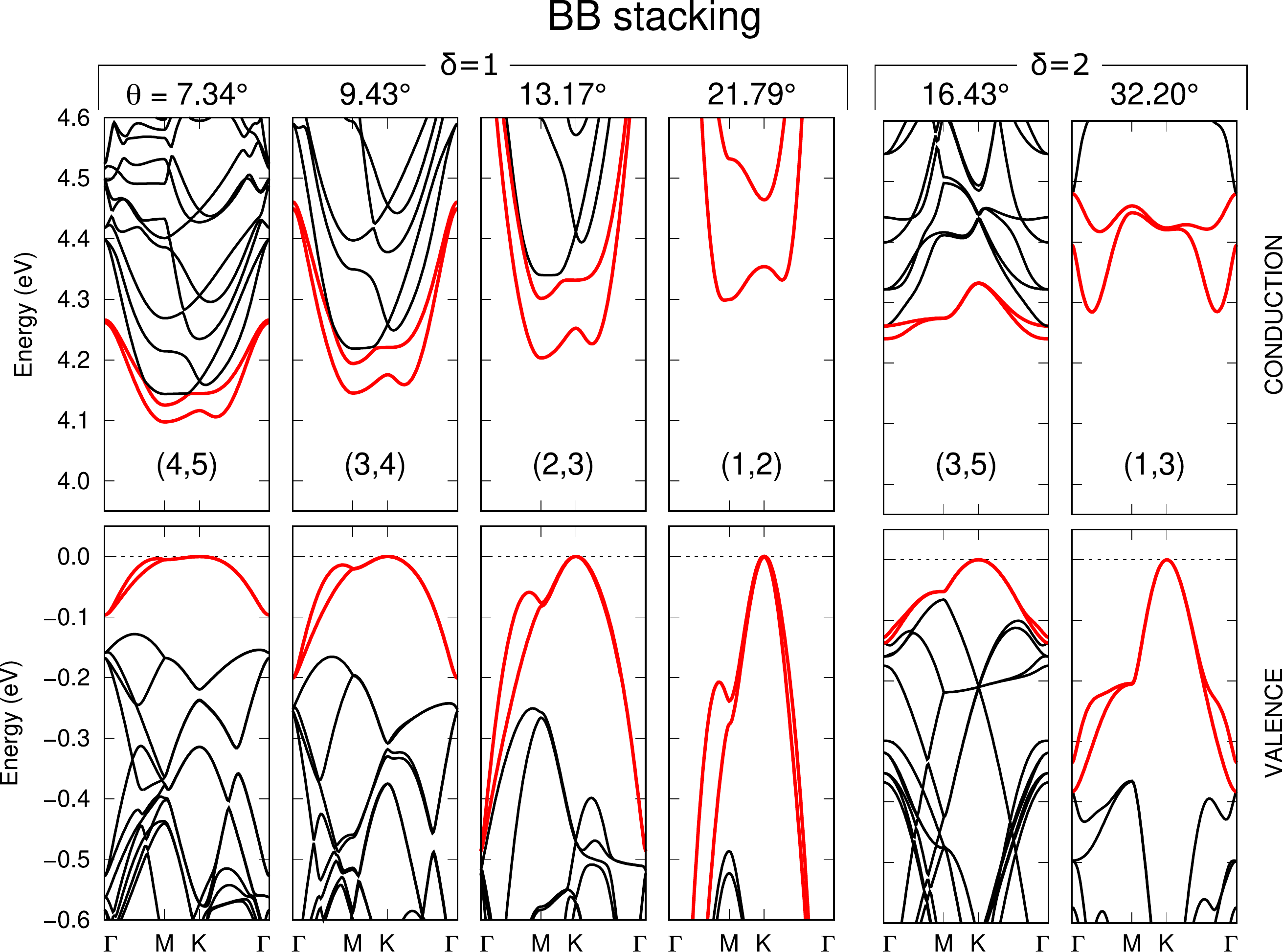}

\vspace{10mm}

\includegraphics[width = 0.49\textwidth]{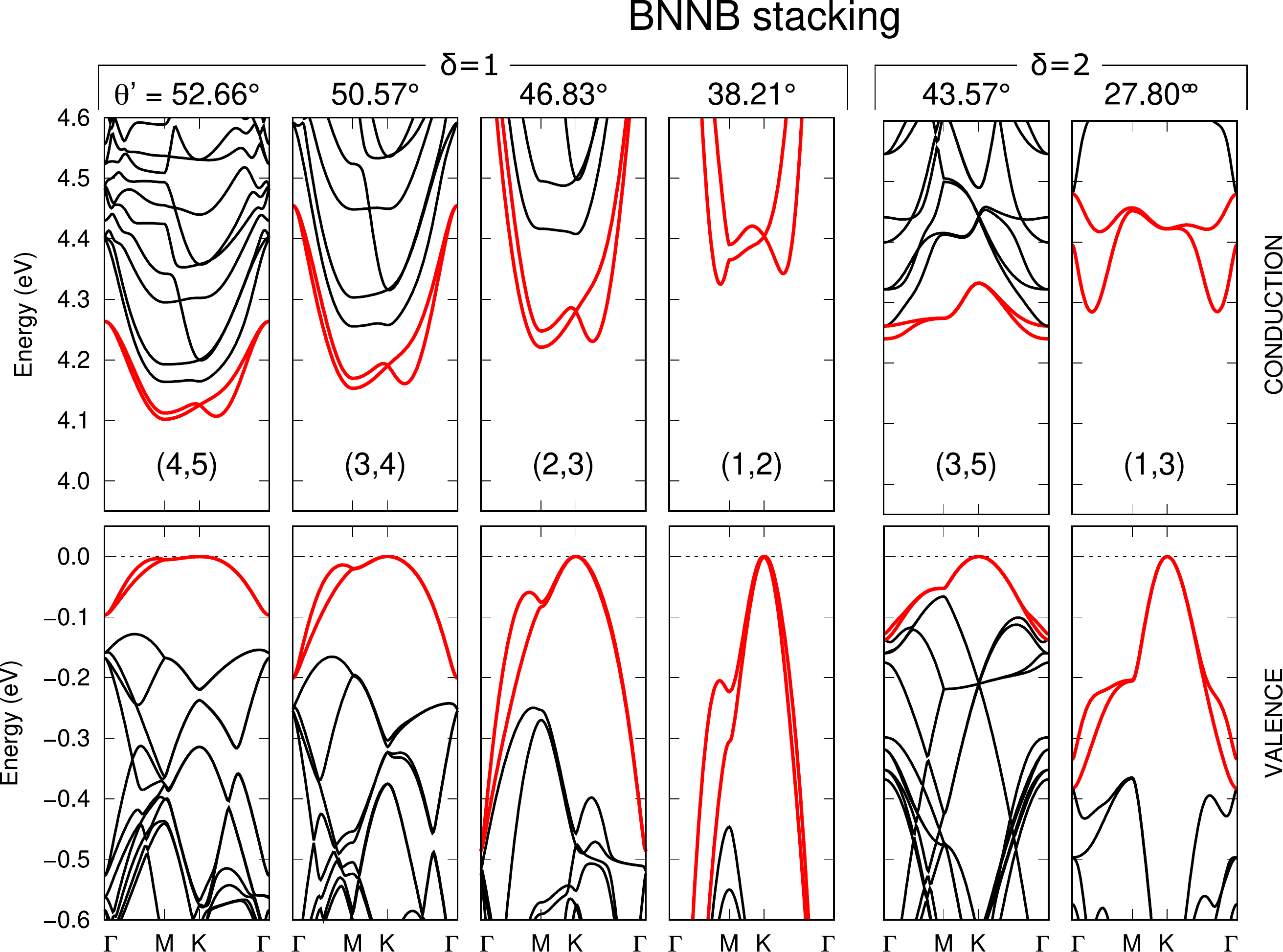}

\caption{Band structure as a function of the twist angle of the BBNN, BB and BNNB stackings from top to bottom.}
\label{fig:band_vs_angles_bis}
\end{figure}

In the main text we give the values of the gapwidth of the five stackings of the (1,3) and (3,5) supercells. 
The values have been extracted from the corresponding band plots, so they refer to gapwidths calculated along specific high symmetry paths in the Brillouin zone. 
In this section we report a more complete mapping of the band structure of the top valence and bottom conduction of the BN stacking, chosen as representative of the bilayers.
In Figure~\ref{fig:band_surfaces} we report the energy surface of the highest occupied states and the lowest unoccupied states in the BN(1,3) and BN(3,5) bilayers. 
With this analysis we demonstrate that the values reported in the main text are meaningful because the bottom of the conduction and the top of the valence fall indeed on the high symmetry lines.

For this analysis we acknowledge F. Paleari who kindly provided us with a dedicated analysis post-processing tool.

\subsection{K: Band structure of the other stackings}
\label{sec:appendix_all-bands}

Here below we report the band plots missing in the main text corresponding to stackings BBNN, BB and BNNB from top to bottom.

\newpage

%

%

\end{document}